\documentclass{aa}  

\usepackage{graphicx}
\usepackage{txfonts}
\usepackage{amssymb}
\usepackage{multirow}
\usepackage{float}
\usepackage{gensymb}
\usepackage{color}
\usepackage{longtable}
\usepackage{rotating}
\usepackage{lscape}
\usepackage[]{hyperref}
\hypersetup{backref=true, pagebackref=true, hyperindex=true, breaklinks=true,colorlinks=true,urlcolor=blue, linkcolor=blue, citecolor=blue,pagecolor=red, bookmarks=true, bookmarksopen=true}

\begin{document}

   \title{Evolution of protoplanetary disks from their taxonomy in scattered light: spirals, rings, cavities, and shadows}


   \author{A.\ Garufi \inst{\ref{Firenze}}
          \and M.\ Benisty \inst{\ref{Santiago_Mixta}, \ref{Grenoble}}
          \and P.\ Pinilla \inst{\ref{Tucson}}
          \and M.\ Tazzari \inst{\ref{Cambridge}}
          \and C.\ Dominik \inst{\ref{Amsterdam}}
          \and C.\ Ginski \inst{\ref{Amsterdam}}
          \and Th.\ Henning \inst{\ref{MPIA}}
          \and Q.\ Kral \inst{\ref{LESIA}}     
          \and \\ M.\ Langlois \inst{\ref{Marseille},\ref{CRAL}}
          \and F.\ M\'{e}nard \inst{\ref{Grenoble}}
          \and T.\ Stolker \inst{\ref{ETH}}
          \and J.\ Szulagyi \inst{\ref{UZH}}
          \and M.\ Villenave \inst{\ref{ESO_Chile}, \ref{Grenoble}} 
          \and G.\ van der Plas \inst{\ref{Grenoble}}
          }

\institute{INAF, Osservatorio Astrofisico di Arcetri, Largo Enrico Fermi 5, I-50125 Firenze, Italy. \label{Firenze}
  \email{agarufi@arcetri.astro.it}  
\and Unidad Mixta Internacional Franco-Chilena de Astronom\'{i}a, CNRS/INSU UMI 3386 and Departamento de Astronom\'{i}a, Universidad de Chile, Casilla 36-D, Santiago, Chile \label{Santiago_Mixta}
 \and Univ. Grenoble Alpes, CNRS, IPAG, 38000 Grenoble, France \label{Grenoble}
\and Department of Astronomy/Steward Observatory, The University of Arizona, 933 North Cherry Avenue, Tucson, AZ 85721, USA \label{Tucson}  
  \and Institute of Astronomy, Madingley Rd, Cambridge, CB3 0HA, UK \label{Cambridge} 
\and Astronomical Institute Anton Pannekoek, University of Amsterdam, PO Box 94249, 1090 GE Amsterdam, The Netherlands. \label{Amsterdam}
 \and Max Planck Institute for Astronomy, K\"{o}nigstuhl 17, 69117 Heidelberg, Germany \label{MPIA} 
  \and LESIA, Observatoire de Paris-Meudon, CNRS, Universit\'{e} Pierre et Marie Curie, Universit\'{e} Paris Didierot, 5 Place Jules Janssen, F-92195 Meudon, France \label{LESIA} 
 \and Institute for Particle Physics and Astrophysics, ETH Zurich, Wolfgang-Pauli-Strasse 27, CH-8093 Zurich, Switzerland. \label{ETH} 
 \and Institute for Computational Science, University of Zurich, Winterthurerstrasse 190, 8057 Zurich, Switzerland \label{UZH}
 \and European Southern Observatory, Alonso de Cordova 3107, Casilla 19001 Vitacura, Santiago 19, Chile \label{ESO_Chile}
    \and Aix Marseille Universit\'{e}, CNRS, LAM, Marseille, France \label{Marseille}
 \and CRAL, UMR 5574, CNRS, Universit\'{e} de Lyon, Ecole Normale Sup\'{e}rieure de Lyon, 46 All\'{e}e d'Italie, F-69364 Lyon Cedex 07, France \label{CRAL}
             }

   \date{Received -; accepted -}

 
  \abstract
   {Dozens of protoplanetary disks have been imaged in scattered light during the last decade.}
   {The variety of brightness, extension, and morphology from this census motivates a taxonomical study of protoplanetary disks in polarimetric light to constrain their evolution and establish the current framework of this type of observation.}
   {We classified 58 disks with available polarimetric observations into six major categories (Ring, Spiral, Giant, Rim, Faint, and Small disks) based on their appearance in scattered light. We re-calculated the stellar and disk properties from the newly available GAIA DR2 and related these properties with the disk categories.}
   {More than half of our sample shows disk substructures. For the remaining sources, the absence of detected features is due to their faintness, their small size, or the disk geometry. Faint disks are typically found around young stars and typically host no cavity. There is a possible dichotomy in the near-infrared (NIR) excess of sources with spiral-disks (high) and ring-disks (low). Like spirals, shadows are associated with a high NIR excess. If we account for the pre-main sequence evolutionary timescale of stars with different mass, spiral arms are likely associated to old disks. We also found a loose, shallow declining trend for the disk dust mass with time.}
   {Protoplanetary disks may form substructures like rings very early in their evolution but their detectability in scattered light is limited to relatively old sources ($\gtrsim$5 Myr) where the recurrently detected disk cavities cause the outer disk to be illuminate. The shallow decrease of disk mass with time might be due to a selection effect, where disks observed thus far in scattered light are typically massive, bright transition disks with longer lifetimes than most disks. Our study points toward spirals and shadows being generated by planets of a fraction of a jupiter  mass to a few jupiter masses in size  that leave their (observed) imprint on both the inner disk near the star and the outer disk cavity. }


\authorrunning{Garufi et al.}

\titlerunning{Evolution of protoplanetary disks from their taxonomy in scattered light}

   \maketitle
%

\section{Introduction}
The most direct observational approach to study planet formation $-$ the direct imaging of forming planets $-$ has thus far been unproductive, with only a handful of young planet candidates being found in the literature \citep[e.g.,][]{Quanz2013a, Sallum2015,Keppler2018,Mueller2018}. On the other hand, the direct imaging of protoplanetary disks has provided several examples of disk sub-structures (e.g., cavities, annular gaps, spiral arms) that are potentially formed by an interaction with embedded (forming) planets. Currently, the paucity of planet detection and the lack of a known evolutionary trend for these substructures are hindering our understanding of planet/disk interaction processes and planet formation.

\begin{figure*}
  \centering
 \includegraphics[width=18cm]{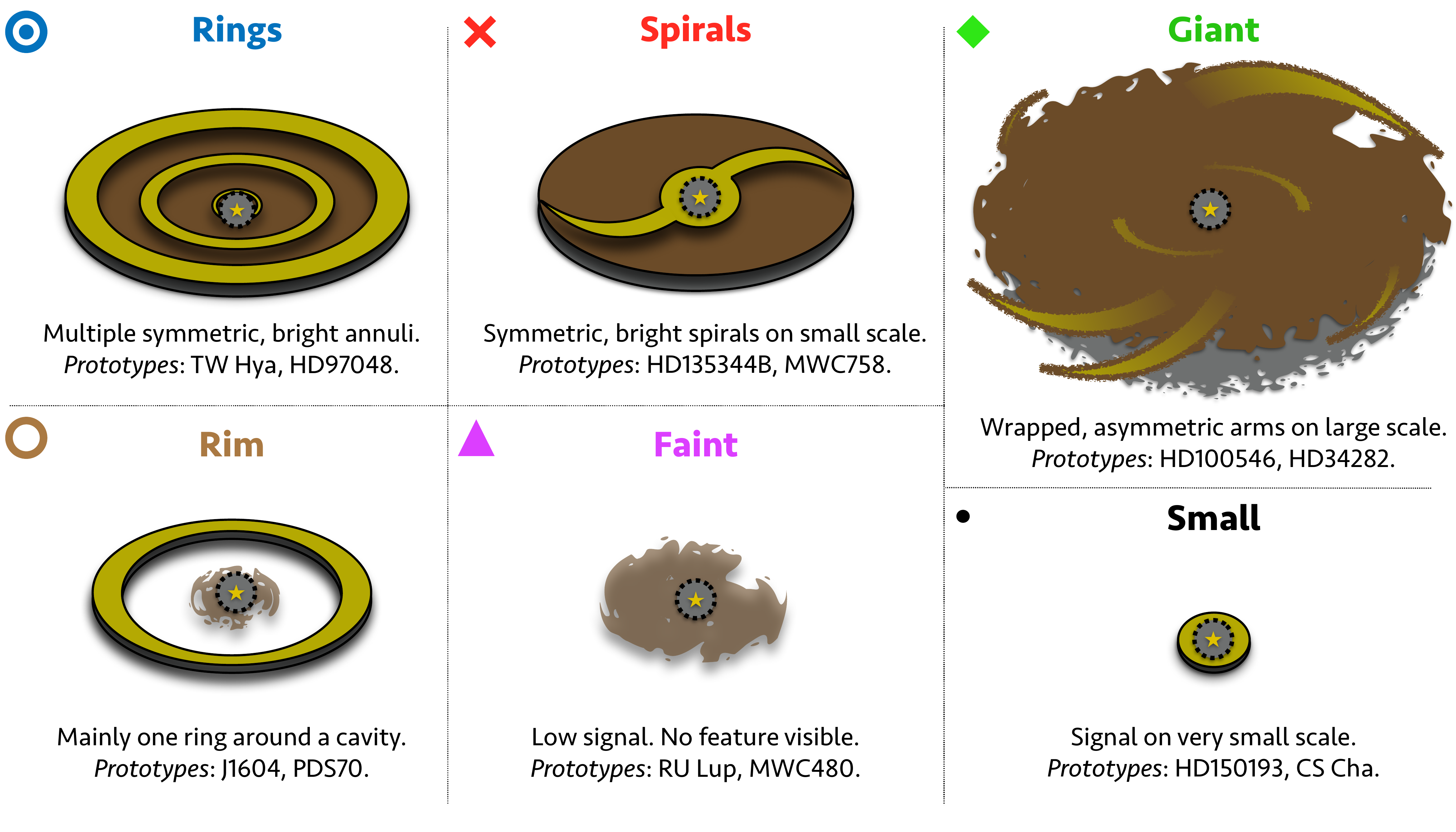}
 \caption{Sketch summarizing the classification of protoplanetary disks in scattered light used in this work (see text for details).}
 \label{Sketch}
 \end{figure*}

Until the advent of ALMA, optical and near-infrared (NIR) imaging of the scattered light from protoplanetary disks was the best method for detecting substructures \citep[see pioneering work by][]{Grady1999, Augereau2001}. Observations at these wavelengths exploit the good angular resolution achieved with single telescopes but suffer from the low contrast of any circumstellar emission in comparison to the star. Much of the current focus is on Polarimetric Differential Imaging (PDI) since this technique guarantees the most efficient removal of the stellar light by separating it from the polarized fraction of the light reflected by the disk \citep[see e.g.,][]{Kuhn2001, Apai2004}. This type of observation traces the small dust grains (from a fraction of a micrometer to a few micrometers in size) at the disk surface and allows one to resolve substructures as small as 5 au from disks at a distance of 150 pc.

Recently, several groups have started a systematic approach to the observation of protoplanetary disks in PDI \citep[see e.g.,][]{Tamura2009, Garufi2017b}. These surveys have revealed an incredibly varied morphology of disks, with the most recurrent feature being an inner disk cavity as large as tens of astronomical units. The existence of these cavities was deduced from the characteristic dip in the near- and mid-IR of their spectral energy distribution (SED) and led to the definition of transition disks \citep{Strom1989} since these objects were initially thought to represent an intermediate stage between protoplanetary and evolved, gas-poor disks. To date, several authors have considered these objects to follow a separate evolutionary path rather than being a particular stage of disk evolution, as derived from, for example, the detection of large cavities already in very young disks \citep{Sheehan2017} and the notion that often these disks are the most massive of all protoplanetary disks \citep{Owen2016}.

Several other sub-structures with no appreciable impact on the SED have been revealed by PDI images. The intriguing discovery of spiral arms \citep[e.g.,][]{Muto2012} and concentric rings \citep[e.g.,][]{Quanz2013b} on scales comparable or slightly larger than the size of our solar system led to the concept of a possible connection with yet unseen planetary companions. The number of disks known to host any of these substructures has now reached the several tens. Nonetheless, an evolutionary framework for the morphology of protoplanetary disks is yet to be established. To a large extent, this lack is to be ascribed to the large uncertainty on the age determination, which in turn partly comes from the uncertainty on the distance to these sources. However, the recent GAIA DR2 \citep{Gaia2018} has alleviated this issue by providing the most accurate measurement of their distance. 

In this study, we catalog the protoplanetary disks with PDI images available from the literature and relate their appearance in scattered light with both the stellar and complementary properties. To do so, we classified all objects into six major categories and studied their appearance in the parameter space defined by the stellar age, mass, luminosity, and variability and by the IR excess and disk mass. We recalculated all properties to ensure homogeneity and to account for the new estimate on the distance from GAIA DR2. The paper is organized as follows. In Sect.\,\ref{Classification} we present the sample and the classification used in this work and in Sect.\,\ref{Recalculation} we describe the recalculation of all properties following GAIA DR2. The main results are presented in Sect.\,\ref{Results} and discussed in Sect.\,\ref{Discussion}. We conclude in Sect.\,\ref{Conclusions}.

\section{Sample and classification} \label{Classification}

The sample studied in this work consists of 58 sources from multiple star-forming regions observed in PDI during the last decade with Subaru/HiCIAO \citep{Tamura2006}, VLT/NACO \citep{Lenzen2003}, GPI \citep{Macintosh2014}, and VLT/SPHERE \citep{Beuzit2008}. We did not consider objects observed by space missions, like for example the Hubble Space Telescope since the inner 1\arcsec \ around these objects was typically not accessible. However, the vast majority of those objects have also been observed by ground telescopes and are therefore included in this work. These 58 stars have spectral types spanning from M1 to B9 and masses from 0.4 M$_{\odot}$ to 3.0 M$_{\odot}$, covering a very wide parameter range of intermediate- and low-mass stars. The complete sample is described in Appendix \ref{Appendix_sample}. We classified all 58 objects into six major categories, based on the appearance (in spatial extent and morphology) of the protoplanetary disk in scattered light. The six major categories proposed are summarized in Fig.\,\ref{Sketch}. 

\textit{Ring disks}. One of the most common features imaged in scattered light are concentric, typically bright rings. Prototypical examples are TW Hya \citep{vanBoekel2017} and HD97048 \citep{Ginski2016}. The disk may show an inner cavity in PDI \citep[e.g., HD169142,][]{Quanz2012} but often it does not show any down to the innermost accessible radius (typically $\approx15$ au).  

\textit{Spiral disks}. A few disks show bright, quasi-symmetric, double spirals on relatively small scale (a few tens of au). The prototypical cases are HD135344B \citep[SAO206462,][]{Muto2012} and HD36112 \citep[MWC758,][]{Grady2013}. Similarly to Ring disks, these disks may (HD135344B) or may not show a cavity (HD36112).

\textit{Giant disks}. We introduce this category to account for those disks showing multiple, asymmetric, relatively faint arm-like structures on large scales. The morphology of the spiral arms detected in for example  HD31293 \citep[AB Aur,][]{Hashimoto2011} and HD142527 \citep{Avenhaus2014a} is in fact clearly different from that mentioned above. In some cases, it is not even obvious whether the arms are wrapped spirals or tenuous, broken rings \citep[HD34282 and HD100546, de Boer et al.\ in prep.][]{Sissa2018}. We leave this case unsolved and refer to these objects as Giant, since these extended, complex arms are always found in disks with very large radial extent ($\gg$100 au). 

\textit{Rim disks}. Some disks do not show any peculiar features besides a bright rim at the outer edge of a disk cavity, like J1604 \citep{Mayama2012} and PDS70 \citep{Hashimoto2012}. Signal within the cavity may be detected and can be strong in some cases \citep[LkCa15,][]{Thalmann2016}. This category only includes transition disks but does not include all transition disks, as many objects of the three previous categories are also transition disks. In other words, the presence of a disk rim is not a characteristic of Rim disks only, but in this category it is the most obvious feature.

\textit{Faint disks}. Not all disks observed in PDI actually show any peculiar feature \citep[e.g., RU Lup,][]{Avenhaus2018}. However, from our experience the non-detection of features is related to the overall faintness of the polarized signal (see Sect.\,\ref{Discussion_faintness} and Appendix \ref{Appendix_faint}). Therefore, it is possible that characteristic substructures are present in the disk but remain undetected \citep[or are barely detected like in AS209,][]{Avenhaus2018} given the current achievable contrast. 

\textit{Small disks}. The detection in PDI of the smallest disks can be hindered by the inner working angle achieved by the current generation of telescopes ($\approx$10$-$20 au at 150 pc). Observations of this type of objects have thus far been mostly succeeded by ALMA \citep[see e.g.,][]{Barenfeld2017} but several attempts in PDI are described by \citet{Garufi2017}, \citet{Ginski2018}, and Dominik et al.\,(in prep.).

We caution that this classification cannot aim at full objectivity and that the categories cannot be perfectly mutually exclusive. The different quality of the various datasets may also have a marginal impact on the classification. In particular, we find a possibly debatable assignment for approximately 20\% of the sample. In view of this, we individually discuss all controversial cases and judge their impact onto the results of this work in Appendix \ref{Appendix_sample}. 

A separate class of objects emerges from the census of imaged disks: the \textit{Inclined disks} \citep[with $i\gtrsim 70\degree$, like T Cha,][]{Pohl2017}. The identification of features in these disks is simply limited by the low viewing angle. However, we do not exclude these sources from the analysis since the  complementary properties that we investigate may also be impacted by the disk orientation.  

An additional element that we consider in this work is the presence of outer stellar companions \citep[e.g., HD150193A,][]{Garufi2014b} and local shadows \citep[HD100453,][]{Wagner2015}. In particular, shadows are seen as more or less narrow dark lanes in scattered-light images where the most realistic interpretation is the occultation of the stellar light by some material close in. Here we only consider shadows with clear indications in PDI \citep[e.g., GG Tau A and HD135344B,][]{Itoh2014,Stolker2016a} and ignore others that have been revealed with other techniques but are not seen in PDI. Furthermore, only dark features that cannot be ascribed to the scattering phase function are considered \citep[i.e., they are not symmetric with the disk orientation like in HD100546 and HD31648,][]{Garufi2016, Stolker2016b, Kusakabe2012}.

\section{Recalculation of stellar and disk properties} \label{Recalculation}
A comprehensive literature is available for the vast majority of the sample. However, our new knowledge of the distance to the sources \citep[GAIA DR2,][]{Gaia2018} affects most of the available estimates. In view of this and to ensure homogeneity, we therefore recalculated several stellar and disk properties for the entire sample.      

\begin{figure*}
  \centering   
  \includegraphics[width=4.85cm]{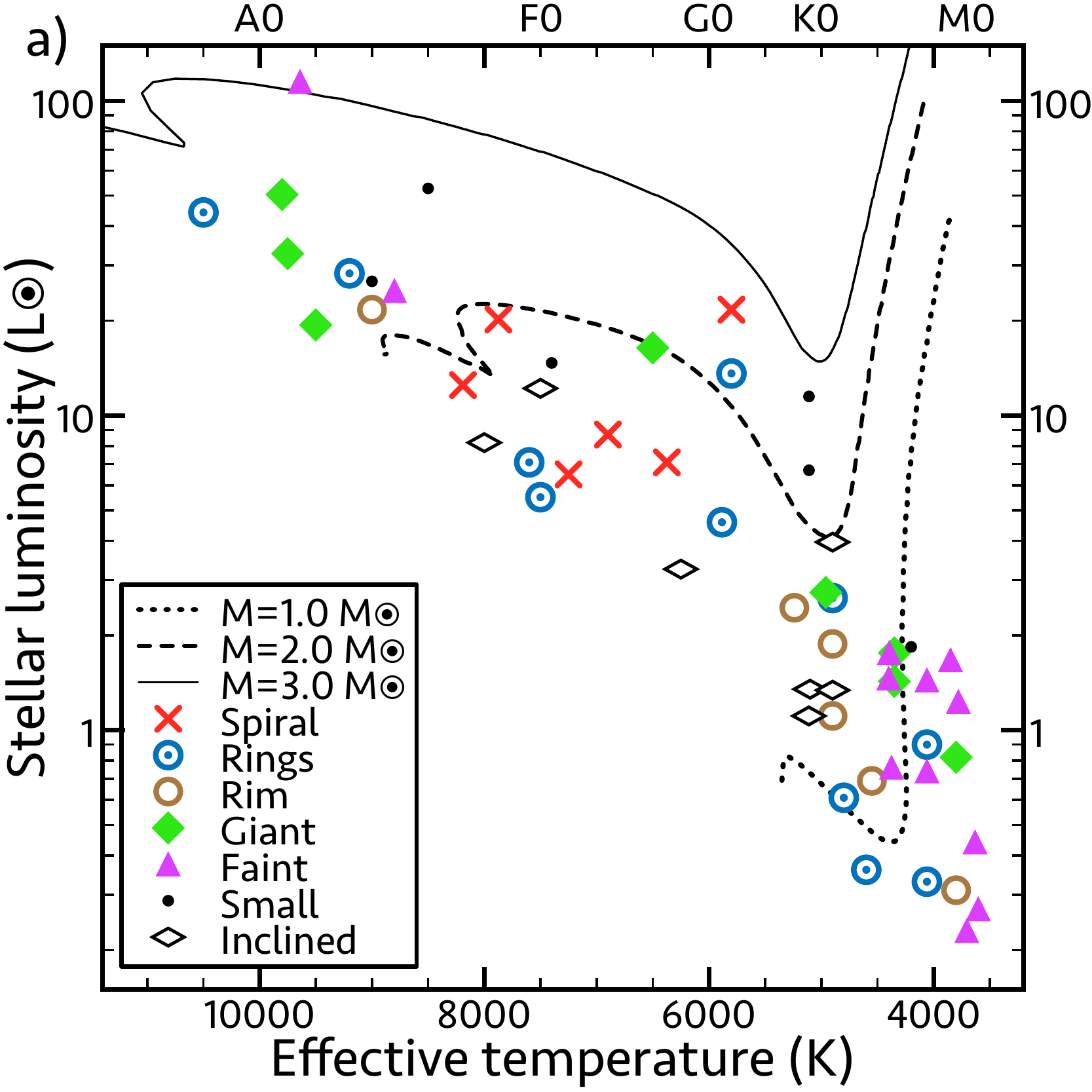}
 \includegraphics[width=6.6cm]{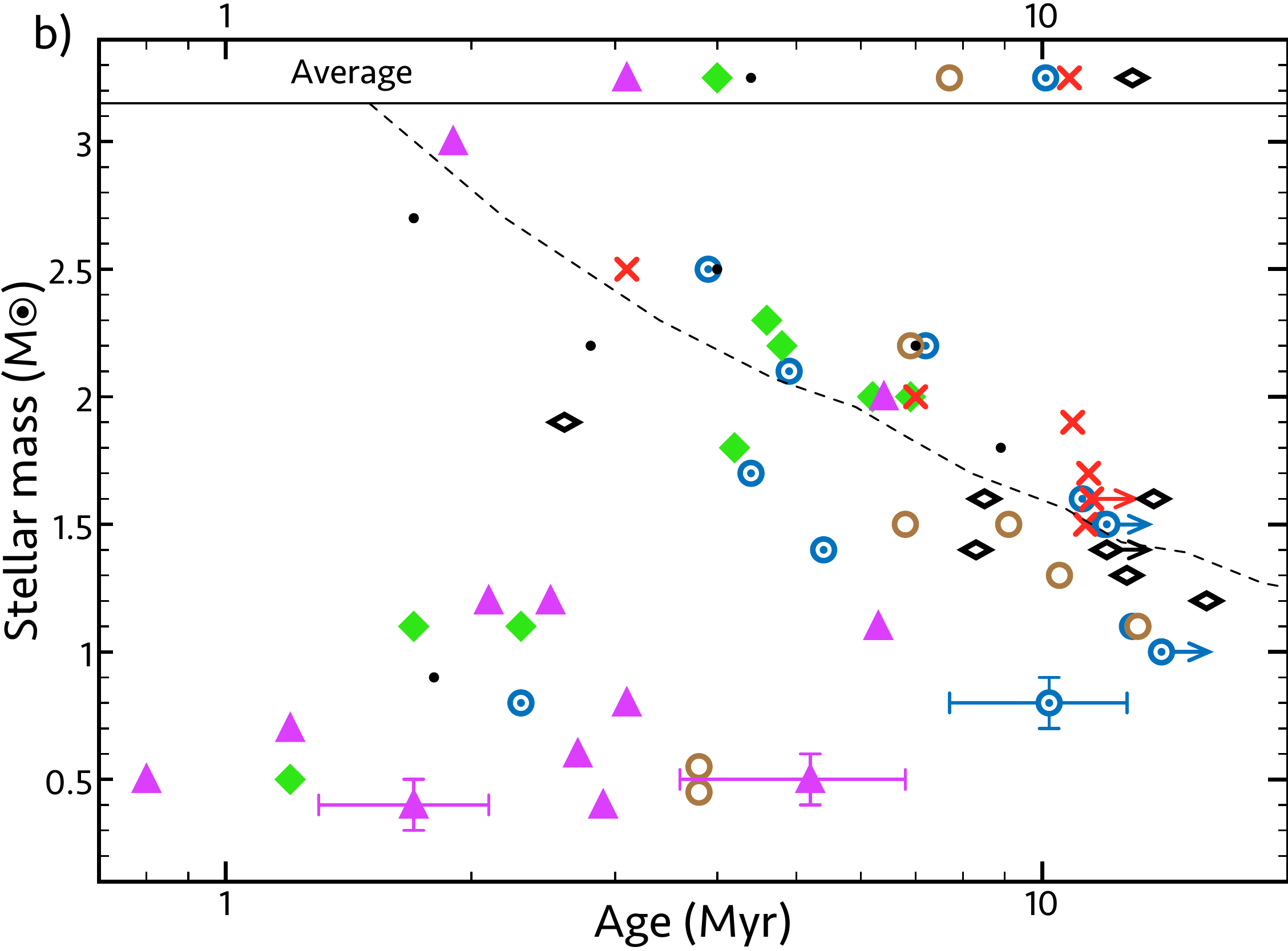}
 \hspace{-0.18cm}
  \includegraphics[width=6.6cm]{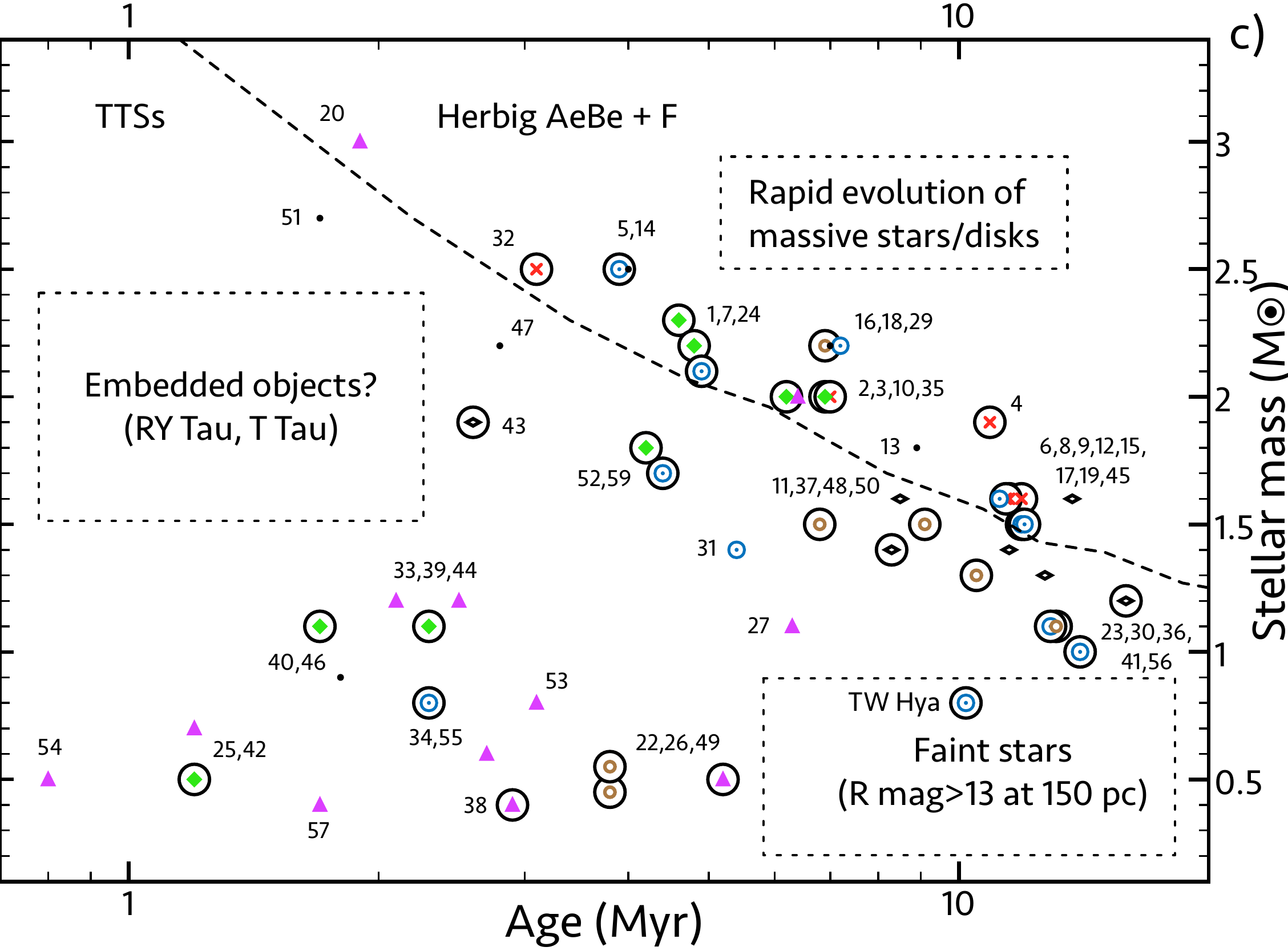}
   \caption{Stellar properties. \textbf{a)}: HR diagram. The three lines are the PMS tracks for the illustrative masses of 1, 2, and 3 $\rm M_{\odot}$. \textbf{b)}: Stellar mass vs. age diagram derived from the PMS tracks. The average of all categories is shown at the top. The dashed line divides Herbig AeBe + F stars from TTSs (G0-type). The three error bars shown are representative of the whole sample at different age intervals. \textbf{c)}: Labeled version of the middle diagram. Circles indicate the presence of a cavity imaged with PDI and/or in the millimeter.}
 \label{Star_properties}
 \end{figure*}

\subsection{New distance from GAIA DR2}
GAIA DR2 provides the most accurate estimate for all the sources studied in this work. In most cases, the previously available distance was within 15\% from the newly available estimate. However, a larger discrepancy is found in a few cases (e.g., Sz111 $-$ from 200 to 158 pc $-$ and LkH$\alpha$330 $-$ from 250 to 311 pc). Interestingly, a few sources have a significantly different distance from GAIA DR1 \citep{Gaia2016} (e.g., HD135344B $-$ from 156 to 136 pc $-$ and V1247 Ori $-$ from 319 to 398 pc). The DR2 uncertainty is on average 1.3 (6.7) pc for our sources within (beyond) 200 pc.

Even though a study of the distance of the various star-forming regions is beyond the scope of this paper, it is worthwhile noticing that:
\begin{itemize}
\item All the Lupus sources (11 in this work) lie in a very narrow range of distance (from 156 to 161 pc).
\item The Taurus sources (9) are systematically further away than what is typically assumed (an average 157 vs 140 pc).
\item Sources from the Sco-Cen association (15) span a large range of distances (from 104 to 166 pc).
\end{itemize}  

\subsection{Stellar properties} \label{Recalculation_stars}
The complete SED {from the B band to 1.3 mm} of each source was built through VizieR\footnote{http://vizier.u-strasbg.fr/viz-bin/VizieR}. In particular, we inspected the temporal evolution of the brightness in the V band through ASAS-3 \citep[from 2000 to 2009,][]{Pojmanski1997} and ASAS-SN \citep[from 2016 to 2018,][]{Shappee2014, Kochanek2017}. This ensured the use of the most representative photometric value and also allowed a quantitative assessment of the stellar variability $\Delta V$. To reject any possible bad measurements, we excluded the $10\%$ highest and lowest values. 

We adopted a PHOENIX model of the stellar photosphere \citep{Hauschildt1999} with the effective temperature $T_{\rm eff}$ and extinction $A_{\rm V}$ taken from the literature, and surface gravity log$(g)=-4.0$. We calculated the stellar luminosity $L_*$ by integrating the photospheric model scaled to the dereddened V magnitude and known distance $d$. Finally, we constrained the stellar age $t$ and mass $M_*$ by comparing the pre-main-sequence (PMS) stellar tracks by \citet{Siess2000} with our newly derived $L_*$ and the literature $T_{\rm eff}$. We adopted solar metallicity, except for a few cases with known depleted metallicity \citep{Folsom2012,Kama2015}. Error bars on $M_*$ and $t$ are derived by propagating the uncertainty on $A_{\rm V}$ (20\%) and distance onto $L_*$ and a conservative $\Delta T_{\rm eff}=200/400$ K for TTSs/Herbig stars. {The uncertainties on $A_{\rm V}$ and $\Delta T_{\rm eff}$ are chosen to account for most different estimates available from the literature while the final error bars do not include the slightly discrepant values of $M_*$ and $t$ found by different evolutionary tracks \citep[see e.g.,][]{Soderblom2014}.} For a few sources older than 10 Myr, the uncertainty is large enough that the adoption of an upper limit is preferable. The adopted literature properties of the entire sample are listed in Table \ref{Sample_literature} while the newly calculated ones are shown in Table \ref{Sample_calculated}.

\subsection{Infrared excess}
From the SED, we obtained a self-consistent measurement of the NIR  and far-IR (FIR) excess {as in \citet{Garufi2017}}. This is done by integrating the flux in excess to the stellar photosphere from 1.2 $\mu$m ({2MASS J band}) to 4.5 $\mu$m ({WISE W2 band}) and from 22 $\mu$m ({WISE W4 band}) to 450 $\mu$m, respectively, {and by normalizing by the stellar luminosity}. The error bars are entirely due to the uncertainty on the $A_{\rm V}$ (20\%) since the error for the photometry is negligible. The calculated values are listed in Table \ref{Sample_calculated}. For some sources, these values were already shown by \citet{Garufi2017}, \citet{Banzatti2018}, and \citet{Avenhaus2018}.

\subsection{Disk mass} \label{Recalculation_diskmass}
We constrained the disk dust mass of each object through the integrated (sub-)millimeter flux. We retrieved the flux $F_{\rm mm}$ at 1.3 mm, and at 0.88 mm when the former was not available (see Table \ref{Sample_literature}) and converted it following
\begin{equation}
M_{\rm dust}=\frac{F_{\rm mm}d^2}{\kappa_{\lambda}B_{\lambda}(T_{\rm dust})}
,\end{equation} 
where $\kappa_{\lambda}$ is the dust opacity and $B_{\lambda}(T_{\rm dust})$ the Planck function at the dust temperature $T_{\rm dust}$. We assumed $\kappa_{\lambda}$ to scale inversely with $\lambda$ from a characteristic value of 10 $\rm cm^2/g$ at the frequency of 1000 GHz \citep{Beckwith1990}. Given the variety of stars of this work, we assumed $T_{\rm dust}=25\ (L_*/L_{\odot})^{1/4}$ K as in \citet{Andrews2013}. The error bars were obtained from the uncertainty on $L_*$, $d$, and $F_{\rm mm}$. Additional uncertainty may come from the possibly different $T_{\rm dust}$ and $\kappa_{\lambda}$ of disks at different evolutionary stages or of different radial extent. In this work, we make use of the normalized $M_{\rm dust}/M_*$, whose values are shown in Table \ref{Sample_calculated}.

\section{Results} \label{Results}
In this section, we show how the different categories of disks proposed in Sect.\,\ref{Classification} are distributed with respect to the environment, stellar, and disk properties calculated in Sect.\,\ref{Recalculation}.

\subsection{Disk features versus stellar properties} \label{Results_stars}
The distribution of the entire sample in the HR diagram is shown in Fig.\,\ref{Star_properties}a. It is clear that with the sources of this work, both stellar luminosity and temperature are rather uniformly covered, reversing the initial trend of a few years ago when hotter stars had primarily been observed. The available sample mostly includes single stars as, to our knowledge, only 12 objects are multiple systems. Of these, five are circumbinary disks (with a triple system in one case, GG Tau A) and eight are circumprimary disks with an outer companion on a large orbit\footnote{The sum yields 13 and not 12 as CS Cha \citep{Ginski2018} belongs to both categories.}. There is no obvious trend with the disk properties, beside the fact that disks are small when they host an outer companion (see Appendix \ref{Appendix_sample}).   

\begin{figure*}
  \centering
 \includegraphics[width=9.2cm]{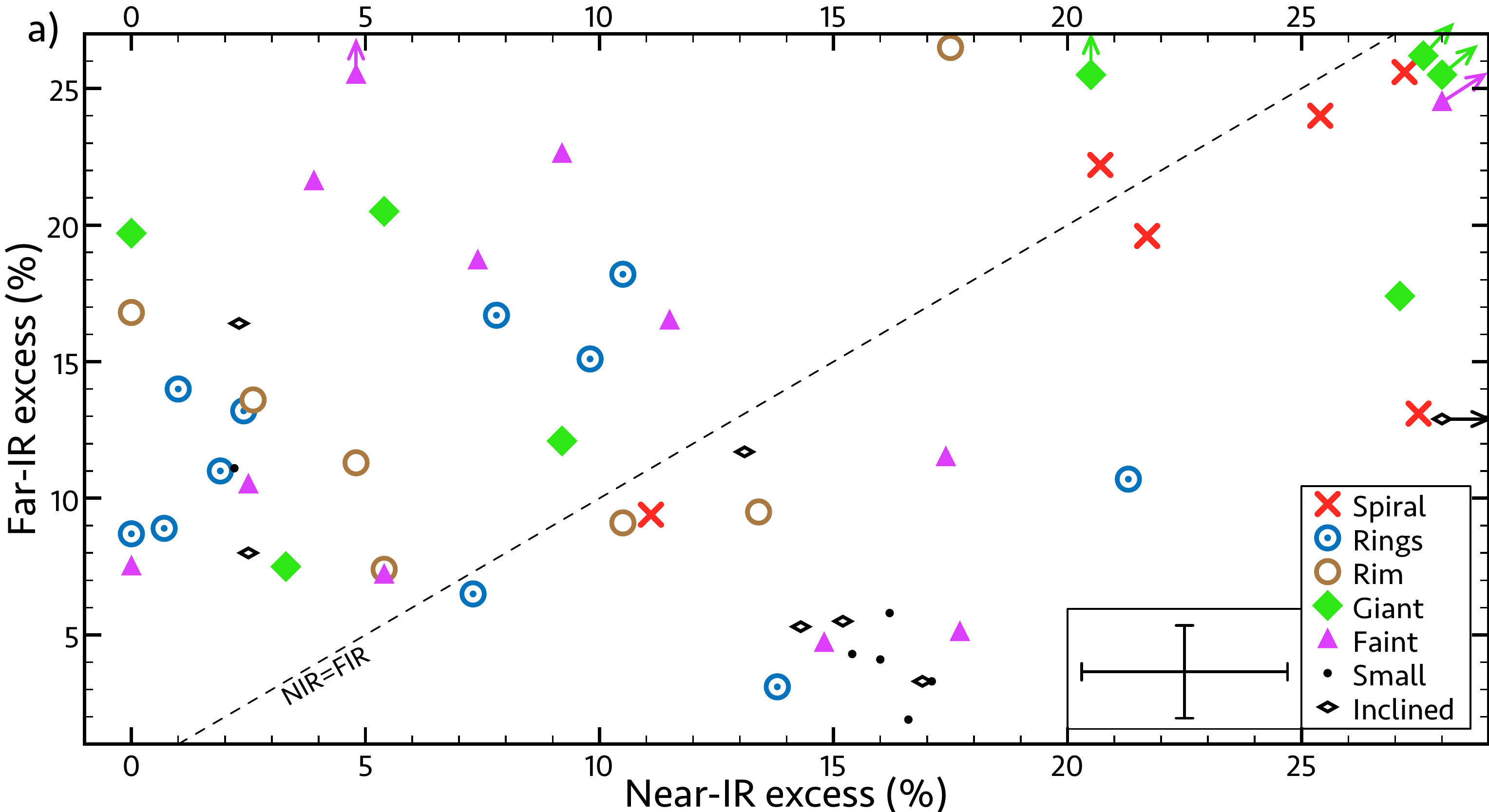}
 \hspace{-0.18cm}
  \includegraphics[width=9.2cm]{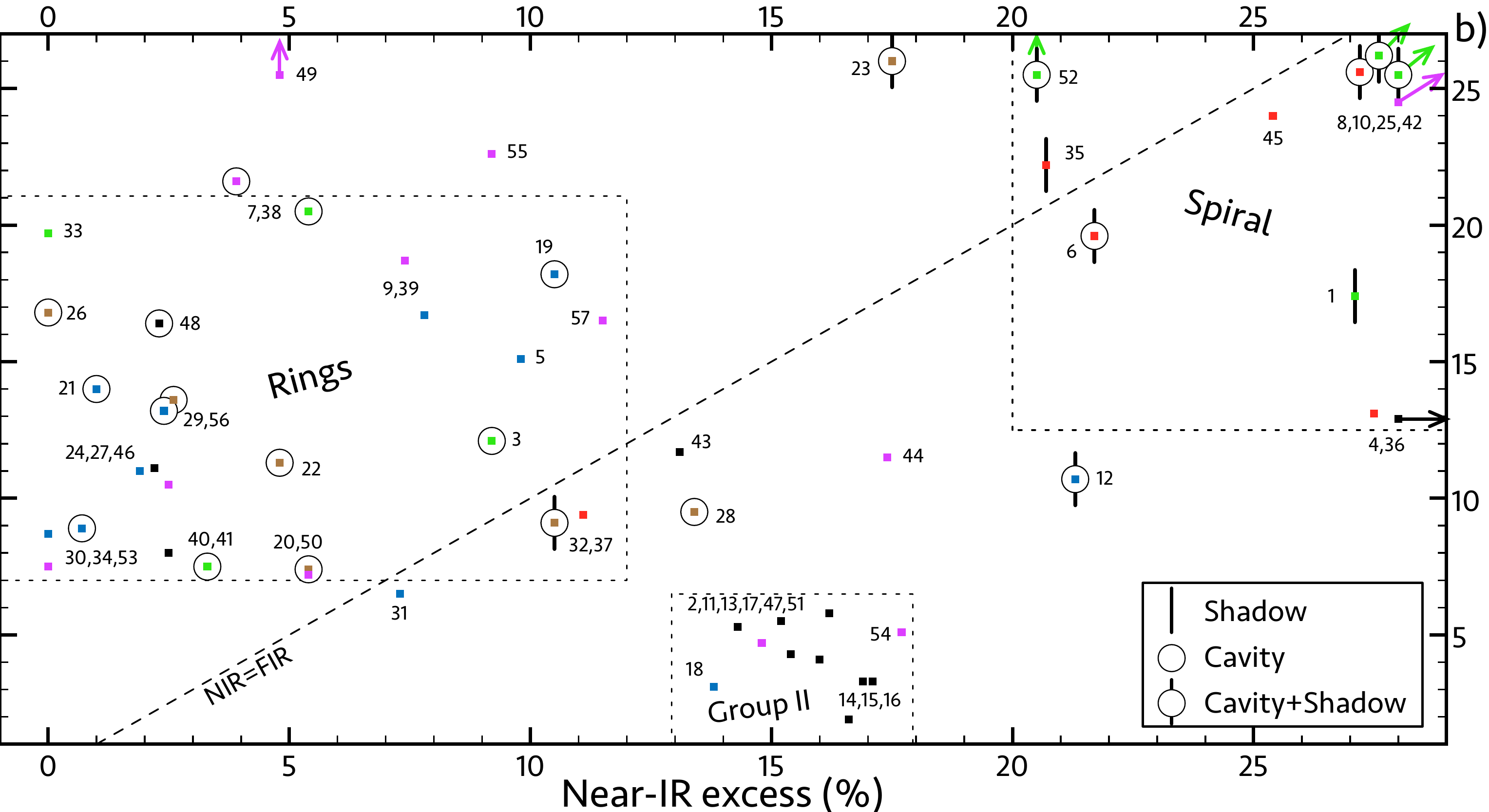}
   \caption{Near-/far-IR excess diagram. \textbf{a)}: Average error bars are shown to the bottom right. The dashed line indicates the equality of the two quantities. Sources with an arrow sit outside of the box. \textbf{b)}: Labeled version of a). The presence of cavities and shadows is indicated by circles and vertical bars.}
 \label{NIR_FIR}
 \end{figure*}

The distribution of disk categories within the HR diagram is uniform, except for two cases: the Spiral disks are thus far only found around earlier-type stars ($T_{\rm eff}>5800$ K) and the Faint disks are primarily associated to late-type stars. As described in Sect.\,\ref{Recalculation_stars}, the comparison between observations and PMS tracks yielded the stellar mass and age shown in Fig.\,\ref{Star_properties}b. Here, it can be seen that the vast majority of Faint disks are young. It is therefore debatable whether the segregation of these disks is intrinsically due to the cold or to the young nature of the host star. At this stage, it should be noted that four Ring disks have similarly low stellar temperature (Fig.\,\ref{Star_properties}a) but only one is of similar age to the Faint disks. Furthermore, the top data point of both diagrams, HD179218, is Faint and the star is hot and young. These aspects may suggest that the accumulation of Faint disks around late-type stars is to be ascribed to their young age, and the observational implication is a segregation around cold stars (since all young, intermediate-mass stars are late-type). To test the significance of this accumulation, we took the distribution {in age} of Faint disks and of the rest of the sample and used the Kolmogorov-Smirnov (KS) two-sided test. The maximum deviation between the cumulative distributions is $\sim$0.6, with a low probability ($\sim$2\%, corresponding to 2.3$\sigma$ significance) that the two samples are similar. This finding is further discussed in Sect.\,\ref{Discussion_faintness}.

From Fig.\,\ref{Star_properties}b, it should also be noted that Spiral disks are always around stars that are old for their given mass (massive stars evolve faster). The KS two-sided test shows a low probability ($\sim$2\%, 2.3$\sigma$ significance) that disks with spirals are of similar age to those without spirals. It should also be noted that stars with Spiral disks are never less massive than 1.5 $\rm M_{\odot}$. Also, Ring disks are typically old with only one source out of twelve being younger than $\sim4$ Myr, while Giant disks are moderately young with none of them in the oldest tertile. These results are discussed in Sect.\,\ref{Discussion_timescale}. Finally, Inclined disks are significantly older than those of all other categories. Unlike the others, this behavior is most likely artificial. For statistical reasons, they should in fact not be different from the others while it is possible that in such geometries a fraction of the stellar radiation is intercepted by the upper disk layer. Consequently, stars will appear fainter, and thus less massive and older (see Fig.\,\ref{Star_properties}a). 

More generally, three deserts clearly appear in Fig.\,\ref{Star_properties}b. One, to the top right, is explained by the shorter evolutionary timescale of {more massive stars, reaching the MS in a few million years. This rapid evolution of the star should therefore reflect a rapid dissipation of the disk making the existence of a disk around an evolved, massive star unlikely.}

Two well-known sources that lie to the left of the diagram (i.e., the young super-Solar stars) are T Tau and RY Tau, namely two bright TTSs associated to extended emission \citep{Csepany2015, Takami2014}. It is possible that stars in this regime are still partly embedded in the natal cloud.

On the other hand, the lack of old sub-solar-mass stars is most likely a sensitivity bias. In fact, we calculated that a {6 Myr-old TTS of 0.5 M$_{\odot}$} at a representative distance of 150 pc has an apparent magnitude in the R band of approximately 13. This value is around the current limit of observability for the current generation of instruments {and restricts the observability of ${<0.5\,{\rm M_{\odot}}}$-mass stars at 150 pc to those older than 2-3 Myr. This limit is even more stringent for extincted and distant sources since it impacts on all sub-solar-mass stars with either $A_{\rm V}>2.0$ or further away than 300 pc (see Table \ref{Limit_Rmag}).} The only object lying in this region of the diagram is TW Hya, which is in fact much closer to us (60 pc) {than all other stars}. 

In Fig.\,\ref{Star_properties}c, the distribution of disk cavities is shown. In particular, we show whether or not disks host a resolved cavity with either PDI or (sub-)mm interferometry. Most of the sources of this work ($\approx$65\%) have a cavity. Sources with no evidence of a cavity are mostly Faint, Small, and Inclined disks. Reasons and implications of this finding are discussed in Sect.\,\ref{Discussion_cavities}. 

Finally, the stellar variability $\Delta V$ estimated as described in Sect.\,\ref{Recalculation_stars} spans from 0.0 to 3.1. The distribution with disk categories is relatively uniform, except for Inclined disks having an average $\Delta V$ significantly higher than the other categories (1.0 vs 0.5) most likely because of occultation of the stellar photosphere by material in the line of sight. Interestingly, no Ring disks have $\Delta V>0.5$ and the average is as low as 0.2 (see Appendix \ref{Appendix_sample}).

\subsection{Disk features versus other disk properties}

\subsubsection{The SED} \label{Results_SED}
The NIR and FIR excess of the entire sample is shown in Fig.\,\ref{NIR_FIR}. Three major clusters of objects are identified. Most sources sit to the left of the equality line, that is, they have a NIR excess numerically equal to or lower than the FIR excess. Eleven sources have intermediate NIR excess and low FIR excess; this region of the diagram identifies the empirical classification of Group II, namely those sources with low far-IR excess \citep[see][and Sect.\,\ref{Discussion_faintness}]{Meeus2001}. Finally, a few sources have both very high NIR and FIR excess.

The diagram shows a hint of segregation for some of the categories depicted in Fig.\,\ref{Sketch}. In fact, the vast majority of Ring disks lie in the major group of objects with intermediate FIR and low NIR excess. On the other hand, five out of six Spiral disks have high NIR excess, which corresponds to a significance probability of 94\% (1.9$\sigma$ significance) of actual segregation from the KS test. Furthermore, all Small disks belong to the Group II cluster. The other categories are spread anywhere in the diagram.

These results suggest that although the physical conditions that contribute to the appearance of faint, rim, and giant disks may be diverse, the dichotomy between spiral and ring disks might be traced back to one specific origin. This element must also have an impact on the appearance of the SED, and in particular on the NIR excess. A similar result was found by \citet{Garufi2017} and \citet{Banzatti2018} and is further discussed in Sect.\,\ref{Discussion_origin}.

\begin{figure*}
  \centering
 \includegraphics[width=9.2cm]{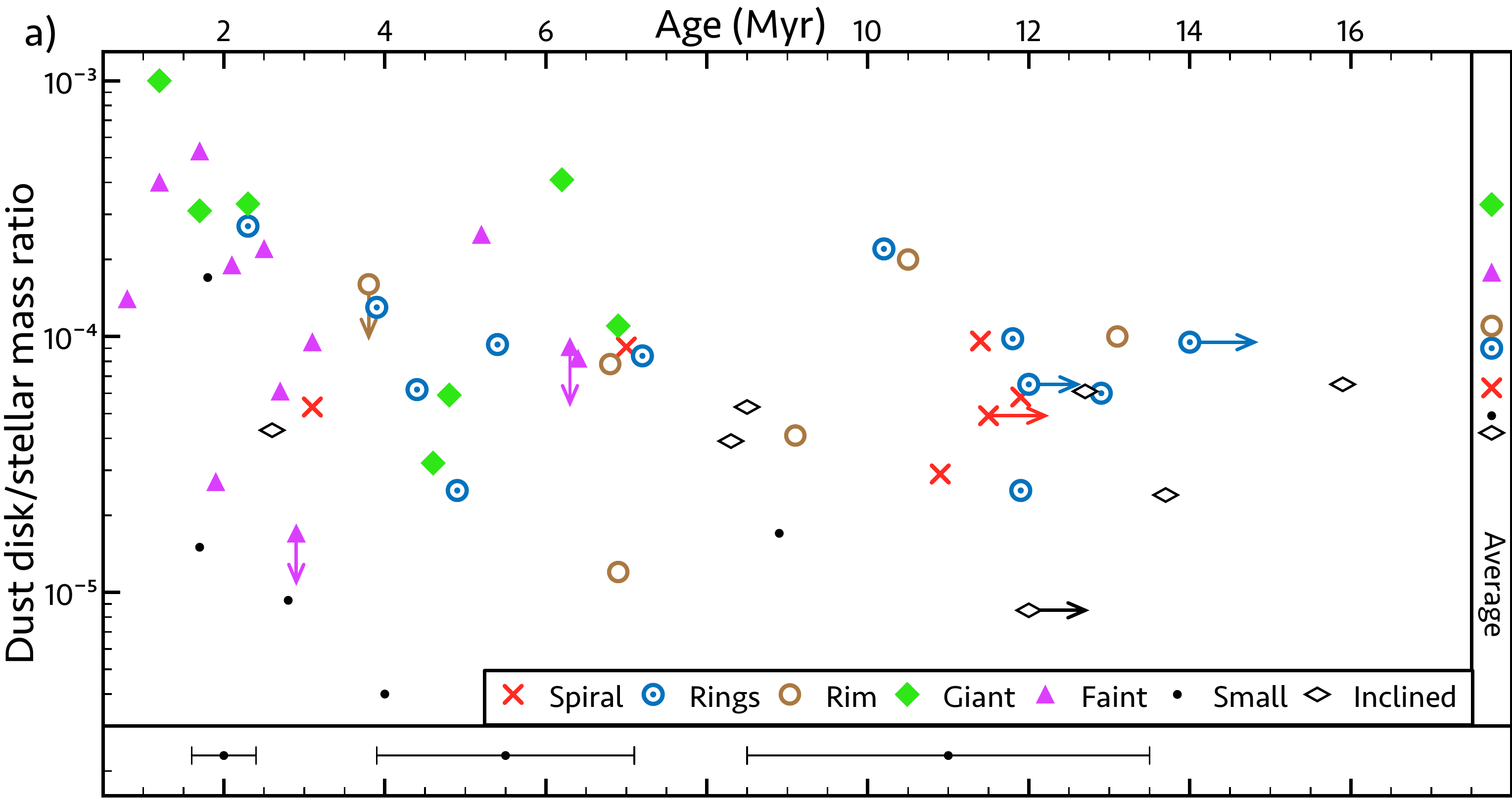}
  \vspace{-0.04cm}
  \hspace{-0.18cm}
  \includegraphics[width=9.2cm]{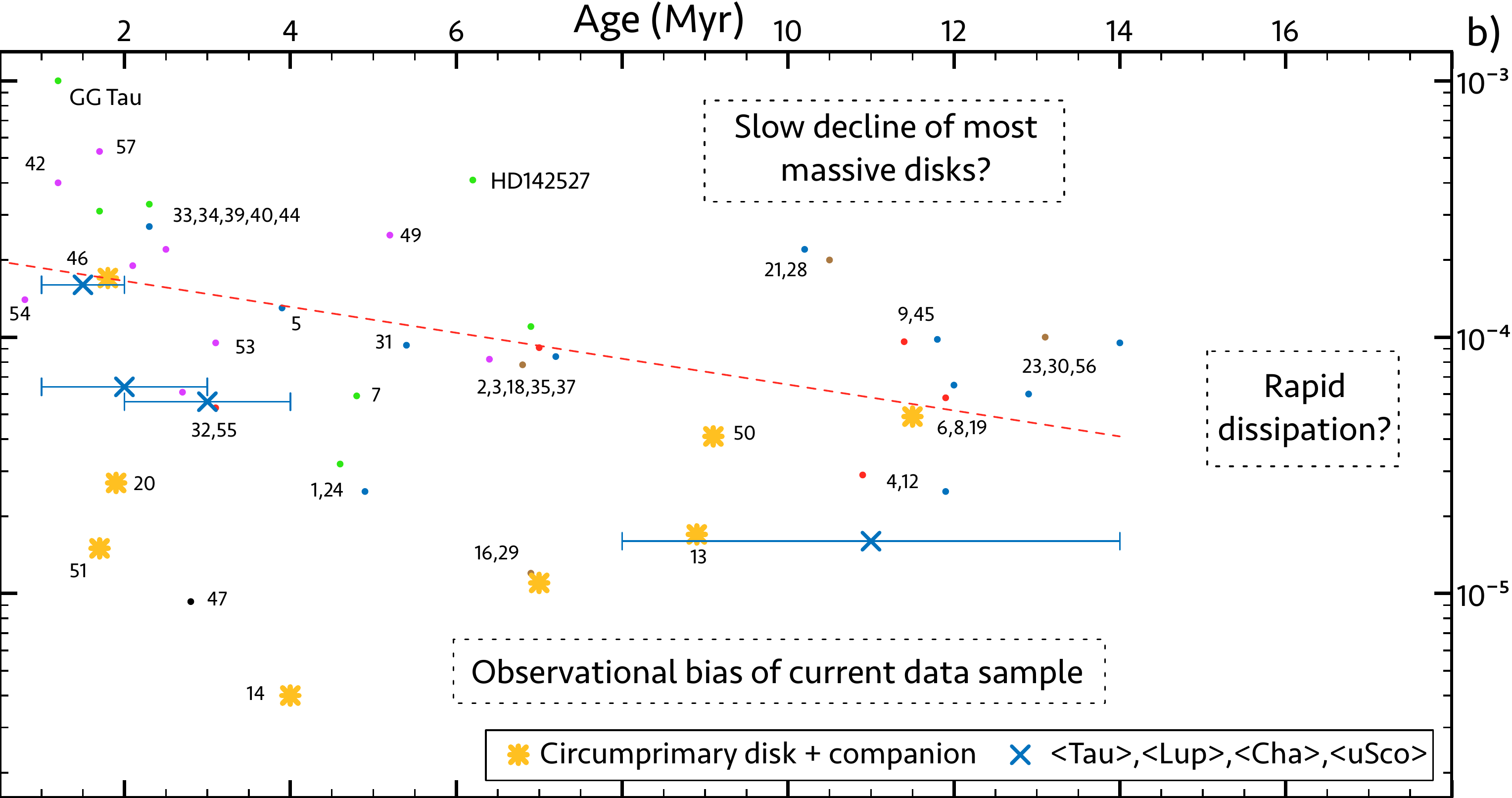}
  \includegraphics[width=9.2cm]{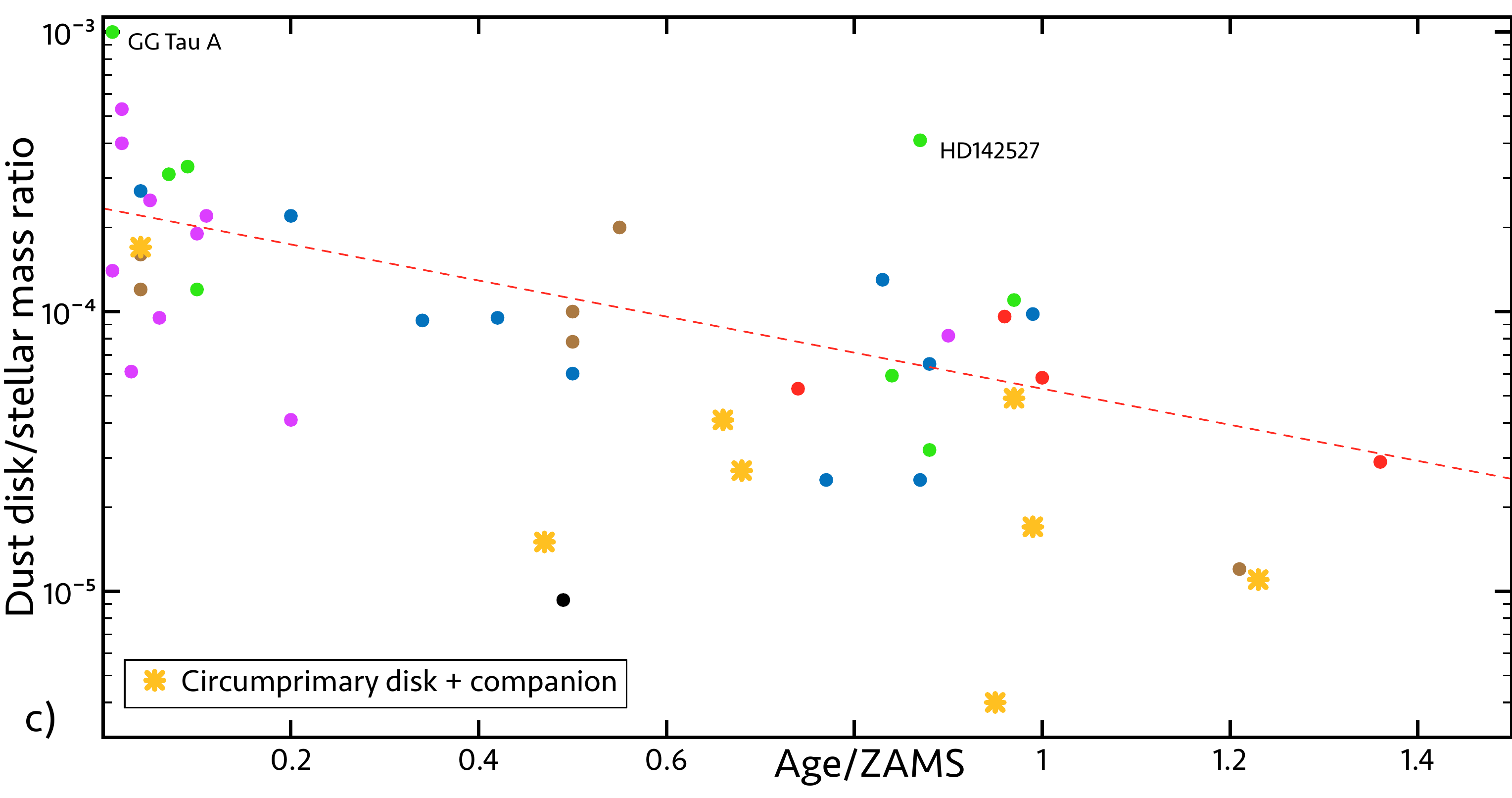}
  \hspace{-0.18cm}
  \includegraphics[width=9.2cm]{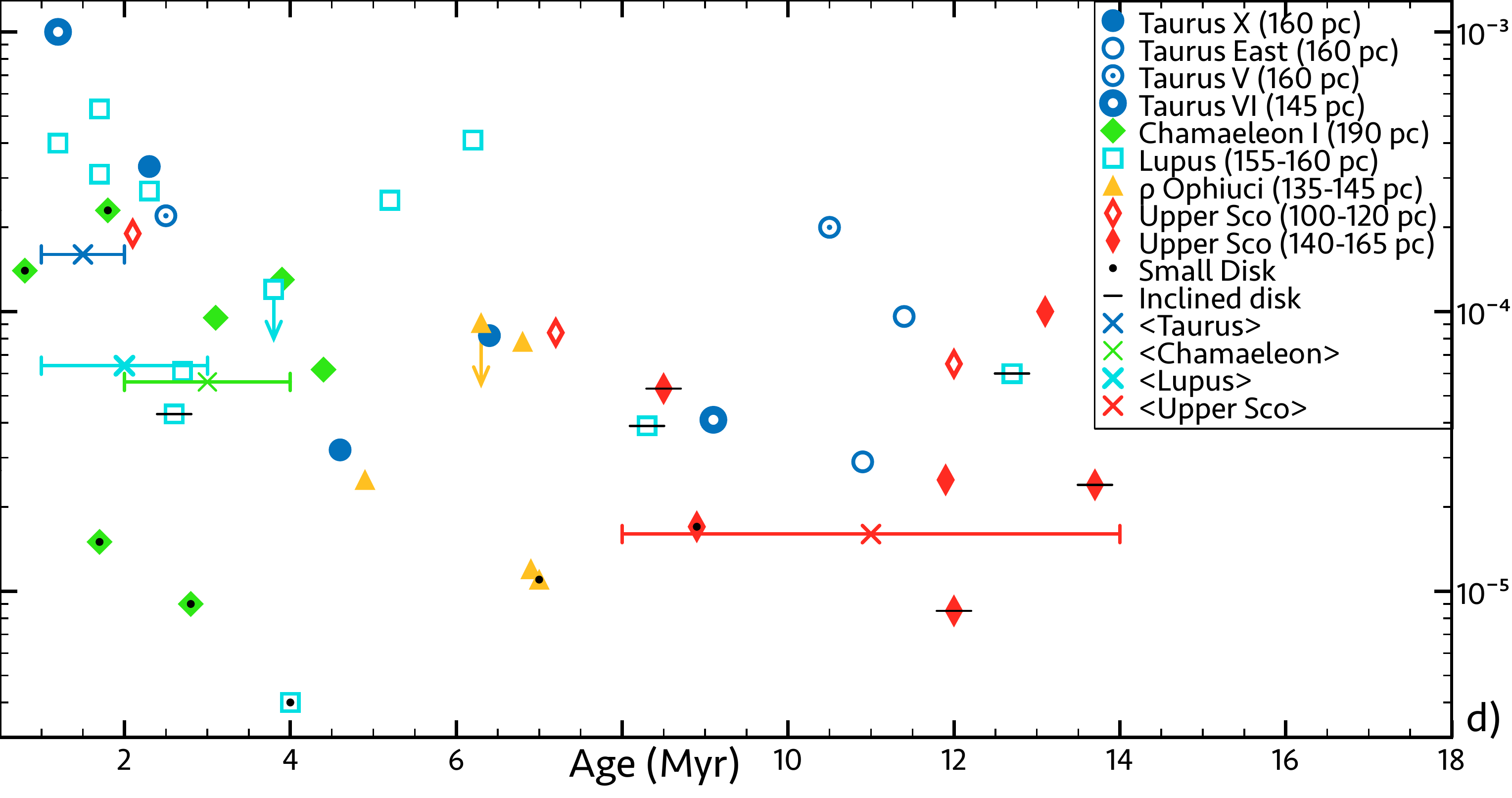}
   \caption{Dust disk mass with time. \textbf{a)}: For the categories of these works. Their average is shown to the right. The error bars in the bottom area are representative of different ages while those on the disk mass are comparable to the symbol size. \textbf{b)}: Labeled version of a). Inclined disks are not included. Disks where an outer stellar companion is present are indicated by the twin dots. The best-fit to our data is shown as a dashed line. The median of Taurus, Lupus, Chamaeleon, and upper Scorpius is shown as crosses. \textbf{c)}: Same as b), with the age being normalized to the ZAMS (see text). \textbf{d)}: Only sources belonging to the main SFRs. Here small disks are indicated by the black dot and inclined disks by the horizontal bar.}
 \label{Dust_time}
 \end{figure*}

The distribution of cavity and shadow across the sample is shown in Fig.\,\ref{NIR_FIR}b. To directly relate cavity and shadow, here we only mark the existence of a cavity if it is detected in PDI (unlike Fig.\,\ref{Star_properties}c). Interestingly, we found that among the sources with a resolved millimeter cavity, as much as $\sim$40\% do not show any cavity in PDI. In Fig.\,\ref{NIR_FIR}b, cavities seem uniformly distributed with the exception of Group II sources, where a cavity is never present. On the other hand, shadows are mostly found in high-NIR sources (with the only exception being DoAr44) and predominantly in Giant and Spiral disks. By performing the KS test in this case, we reveal that there is also a very low probability that disks with shadows have a similar NIR excess to the rest of the sample ($\sim$6\%). In particular, it should be noted that for all the sources with NIR $>15\%$ the presence of a cavity in scattered light implies the presence of a shadow. This trend is discussed in Sect.\,\ref{Discussion_origin}.

\subsubsection{Disk masses}
The dusty disk/stellar mass ratios obtained in Sect.\,\ref{Recalculation_diskmass} are related to the stellar age in Fig.\,\ref{Dust_time}a. A clear yet loose trend is visible in the diagram with several young sources having a mass ratio between $10^{-3}$ and $10^{-4}$ and all old sources having one $\lesssim10^{-4}$. Any quantitative considerations on the trend suffer from the large uncertainties on the age (in particular for $t\gtrsim5$ Myr). The Pearson correlation coefficient for the relation is $-0.37$, indicating a weak correlation between the two quantities. 

Interestingly, a large fraction of the young, Faint disks discussed in Sect.\,\ref{Results_stars} are among the most massive disks of the entire sample. A few others have instead comparably low mass to the Small disks suggesting that diverse geometries contribute to the faintness in scattered light. Spiral disks have slightly lower dust mass than the other categories with features. Finally, Inclined disks all lie in the lowest half of the distribution, which is contradictory from statistical considerations. This may indicate that dust masses of disks seen almost edge-on may be significantly underestimated {because of marginally, optically thick emission even at millimeter wavelengths}. 

In Fig.\,\ref{Dust_time}b, we show how the sources of this work relate to the median from different star-forming regions (SFRs). We extracted the best-fit to our data (red, dashed line) excluding non-detections in the millimeter and Inclined disks for which the age determination is possibly biased (see Sect.\,\ref{Results_stars}) and compared it to the median (blue crosses) found for Taurus, Lupus, Chamaeleon, and Upper Scorpius (Villenave et al. in prep.). Clearly, this line lies above the SFRs indicating that disks observed thus far in scattered light represent the upper tail in the mass distribution. The few sources of our sample lying below the median of SFRs are typically Small disks, with an outer stellar companion that is likely truncating the circumprimary disk. Given the overall trend, the absence of disks around stars older than 15$-$20 Myr clearly implies that a more rapid dissipation eventually occurs.

In Fig.\,\ref{Dust_time}c, we also compare the disk mass with the age normalized to the time needed to a star with the respective mass to reach the zero age main sequence \citep[ZAMS, from][]{Siess2000}. In this case, the majority of sources seem to align on the trend, with the exception of the Small, truncated disks and the exceptionally massive GG Tau and HD142527. More quantitatively, the correlation is stronger as the Pearson coefficient is $-0.50$. This is likely an indication that the dust dissipation is more easily observable when accounting for the different evolution timescales of stars with a different mass.

\subsection{Disk features versus the environment}
Most of the sources observed thus far in scattered light belong to a specific SFR. The only exception is, to our knowledge, HD179218, while TW Hya and V4046 Sgr are part of stellar groups devoid of any other protoplanetary disks. A few sources are the only representative of more or less large regions, like for example IRAS 08267-3336 (Gum Nebula), RX J1852.3-3700 (Corona Australis), and V1247 Ori (Orionis OB1). Three sources belong to $\epsilon$ Cha (PDS 66, T Cha, DZ Cha). 

Much larger samples are instead provided by Taurus, Chamaeleon, and the large associations of Lupus, $\rho$ Ophiuci, and Scorpius-Centaurus. With the distance of all objects now available,  studying the distribution of more complete samples within these regions is certainly worthwhile. Within the framework of this work however, we focus on the four dozen objects with resolved scattered-light observations. 

\textit{Taurus.} Our nine sources in Taurus belong to four different sub-regions (VI, V, X, and the eastern edge) and this is reflected in very different ages and therefore disk masses (see Fig.\,\ref{Dust_time}d). Given this, the comparison to the average disk mass is probably meaningless. In PDI, the Taurus sources imaged thus far already show all sorts of structures but they are typically extended (i.e., no Small disk belongs to Taurus). 

\textit{Chamaeleon.} Unlike Taurus, all Chamaeleon sources (7) are very tightly distributed (all within 20 pc of relative distance). Their age is in fact limited to $\approx4$ Myr. Interestingly, four out of seven disks appear Small. The only prominent disk thus far observed in scattered light is HD97048.

\textit{Lupus.} All Lupus sources (12)\footnote{In this work, we consider HD142527 as Lupus source. Even though it is typically associated to the Upper Cen association, its new distance makes it closer to Lupus} are relatively close in space and age (see Fig.\,\ref{Dust_time}d). The only sources that in our calculation appear much older than the rest of the sub-sample are two Inclined disks, MY Lup and IRAS 16051-3820. This is possibly additional evidence that the age determination from inclined disks is problematic (see also Sect.\,\ref{Results_stars}). From the comparison with the average disk mass for Lupus, a very strong bias toward massive disks emerges. Nonetheless, many disks are Faint in scattered light (whereas others $-$ like IM Lup and RX J1615 $-$ are very prominent).

\textit{Upper Sco and $\rho$ Oph.} On the sky plane, $\rho$ Oph is embedded in Upper Sco. The 3D map now shows that its sources are intrinsically very close to some sources in Upper Sco\footnote{In this work, we consider HD150193 as $\rho$ Oph source. Even though it is typically associated to the Upper Sco association, its new distance makes it closer to $\rho$ Oph, namely those at 140-165 pc}. From Fig.\,\ref{Dust_time}d, the Upper Sco sources appear as an ideal prosecution in time of $\rho$ Oph. Other Upper Sco sources at smaller distance and the Upper Cen sources instead span a much larger interval of age and disk mass.

Figures \ref{Dust_time}d also shows how the trend for the disk mass with age in these regions is possibly less dispersed and less shallow than the main one. In fact, drawing only sources that do not belong to these regions results in no trend.

\section{Discussion} \label{Discussion}
In this section, we discuss the implications of the primary results listed below that emerged from the analysis of Sect.\,\ref{Results} and that only concern the 58 sources of this work and are found with a $\sim$2$\sigma$ significance:
\begin{itemize}
\item Faint disks are young.
\item Spiral disks are old, being at the end of their PMS evolution. They have high NIR excess and moderate disk mass.
\item Ring disks have low NIR excess and stellar variability, and no outer stellar companion or shadow.
\item Shadows are seen in sources with high NIR.
\item There is a loose, shallow, declining trend for the dust disk mass with age (which is slightly stronger when we account for the evolutionary timescales of stars with different mass).
\item Current estimates of age and mass of Inclined disks may be incorrect
\item We are currently biased toward old super-solar and young sub-solar stars, as well as toward massive disks.
\end{itemize} 

\subsection{Timescale for the formation of substructures} \label{Discussion_timescale}
Addressing the timescale for the formation of disk features is also pivotal in understanding their origin. As an example, the ALMA image of HL Tau \citep{ALMA2015} shows multiple rings that might be inconsistent with planet/disk interaction in the core accretion scenario \citep{Pollack1996}, given the young stellar age. When studying the temporal evolution of any disk feature in scattered light, we must keep in mind that super-solar-mass stars observed thus far are primarily old whereas sub-solar stars are primarily young (see Fig.\,\ref{Star_properties}b).  

In this work, we show that rings are typically imaged in scattered light around older stars. This notion by itself would suggest that these structures are formed a few million years after the stellar formation. However, we also show that many young disks are actually faint in scattered light and this may hinder the detection of disk features. The prototypical example of this behavior is AS 209 where the disk clearly shows multiple rings in the sub-millimeter \citep{Fedele2018} that are only barely detected in scattered light \citep{Avenhaus2018}. Therefore, the accumulation of Ring disks around older stars may only be due to the global faintness of young objects in scattered light (discussed in Sect.\,\ref{Discussion_faintness}) and rings may well form early ($\ll$2 Myr) but only become detectable later in the disk lifetime ($\gtrsim$2 Myr).

In principle, the same considerations may apply to spirals. In fact, we find that the six stars hosting spiral-disks are all at a very late stage of their PMS evolution. In absolute terms, their age varies from $\sim$3 Myr to $\sim$12 Myr depending on their mass. In relative terms, these ages correspond to $>$80\% of their PMS lifetime. Unlike Ring disks, we found no source at an earlier evolutionary stage suggesting that the trend may be due to an actual late formation for spiral arms. This scenario also provides a possible explanation for the lack of spiral disks around TTSs. In fact,  very few old ($>$5 Myr) low-mass stars have been observed in scattered light (see Fig.\,\ref{Star_properties}b).

\subsection{Origins of substructures} \label{Discussion_origin}
Since the start of protoplanetary disk imaging, it has been clear that disks often show substructures. From this work, it turns out that when the disk size, brightness, and orientation allow it, such features are always detected. In fact, the only sources showing no extended arms, spirals, or rings are the inclined disks (because of the geometry), the faint disks (because of the sensitivity of the observations), and the small disks (because of the angular resolution). Clearly, the first category does not represent an observational bias as we can assume that inclined disks appear identical to the others. As discussed in Sects.\,\ref{Discussion_timescale} and \ref{Discussion_faintness}, in faint disks the same structures may be present but remain elusive. 

As for small disks, we cannot conclude anything about the existence of substructures. In particular, it would be interesting to investigate whether the extended structures that are recurrently observed in giant disks are also present in small disks, making them a scaled-down version of the large disks. This work (and the current generation of telescopes) does not answer this question. Nonetheless, it must be noted that the NIR excess of these objects is typically within a narrow range of values \citep[see Fig.\,\ref{NIR_FIR}a and][]{Banzatti2018} suggesting that the physical conditions of the inner regions are more uniform than for extended disks.  

\subsubsection{Cavities} \label{Discussion_cavities}
The high occurrence of cavities in the sample of imaged disks is well known. Any considerations on the distribution of disk cavities within this sample may be biased by the abnormally large fraction of transition disks {in current PDI samples}. As mentioned in Sect.\,\ref{Results_stars}, $\sim$65\% of our sources show a cavity and this is a lower limit to the fraction of transition disks since some cavities may be too small to be resolved \citep[see e.g.,][]{Menu2015}. Splitting the sample into young and old sources with threshold at 4 Myr yields $\gtrsim$50\% and $\gtrsim$70\%, respectively. These numbers must be compared to those obtained photometrically from large samples like the 8\% and 46\% {fraction of old and young transition disks,} respectively, by \citet{Balog2016} or the generally accepted $\sim$10\% \citep[see review by][]{Owen2016}. Therefore, the current sample is clearly biased toward transition disks and any estimate on the temporal evolution of disk cavities may be partial.  

Focussing on our sample, a disk cavity has been imaged in scattered light and/or in the \mbox{(sub-)millimeter} for all the Rim (by definition), Spiral, Giant, and for most of the Ring disks. A significant exception within the Ring disks is HD163296 \citep{MuroArena2018}. On the other hand, a disk cavity is rarely imaged in the Faint, Small, and Inclined disks. For Faint disks the absence of a large cavity could be the origin of their faintness (see Sect.\,\ref{Discussion_faintness}), while for Small disks it is an obvious consequence of their nature, and for Inclined disks it is an observational bias. In fact, in scattered light an inclined disk likely hides the cavity. In principle, (sub-)millimeter images should always reveal a cavity but it is not obvious whether all disks are actually optically thin at those wavelengths.

An important element to understand the origins of disk cavity is the large number of disks with a cavity detected in the \mbox{(sub-)millimeter} but not in PDI ($\sim$40\%, see Sect.\,\ref{Results_SED}). Within the remaining $\sim$60\%, approximately half (i.e., $\sim$30\%) show a smaller cavity in PDI than in the millimeter \citep[e.g.,][]{Garufi2013}, resulting in $\sim$70\% of the millimeter-cavities being smaller in the NIR. This morphology is a natural result of dust filtration. In particular, the spatial segregation of small and large grains is indicative of dust trapping at the outer edge of a gap carved in the gas surface density by a massive planet \citep{Pinilla2012, deJuanOvelar2013}. Other mechanisms that lead to particle trapping, such as the outer edge of a dead zone, do not lead to this segregation resulting in cavities of similar size at short and long wavelengths \citep{Pinilla2016a}.

\subsubsection{Shadows} \label{Discussion_shadows}
An important finding of this work is that sources showing a shadow in the outer disk systematically have a high NIR excess \citep[see Fig.\,\ref{NIR_FIR}b and also][]{Banzatti2017}. In particular, when a high-NIR source shows a disk cavity in scattered light it also shows a shadow. For disks without a scattered-light cavity, like for example HD36112, shadows may be present at a smaller scale than what is currently accessible with PDI observations. 

The most plausible explanation for shadows in the outer disk is the existence of a dust belt at a  distance to the star of less than 1au. This belt must be sufficiently misaligned with respect to the outer disk to allow the projection of azimuthally confined shadows \citep{Marino2015, Benisty2017, Min2017}. In turn, the misalignment is best explained by the presence of a massive planet or a low-mass star on an inclined orbit \citep{Facchini2013}. Our finding that the presence of shadows corresponds to a high NIR excess implies that this inner disk must also be vertically extended. In fact, all possible scenarios for an increased NIR \citep[e.g., a magnetically supported disk atmosphere or the presence of warps induced by a companion,][]{Flock2017, Owen2017} invoke an increased vertical scale height of the dusty structures.

Therefore, massive planetary companions within the disk cavity can explain both the misalignment and the vertical extent of inner disk structure that are responsible for shadows and high NIR. However, in the case of a massive planet being the origin of the cavity, the NIR excess is expected to decrease with time. In particular, disks older than 3$-$5 Myr that host a very massive planet ($\gtrsim$5 M$\rm _{Jup}$) {on a circular orbit} should not show any NIR excess, since all the dust is filtered in the outer disk and the optically thick dust belt near the star cannot be replenished \citep{Pinilla2016b}. Conversely, the NIR in our old sources is systematically high. This implies that the planetary scenario is consistent with our observations only if the putative companion has a mass that is between a fraction of a Jupiter mass and a few Jupiter masses. 

\subsubsection{Spirals} \label{Discussion_spirals}
Similarly to shadows, spiral arms are likely associated with a high NIR excess (see Fig.\,\ref{NIR_FIR}). The consequence of this double connection is the known association between spirals and shadows \citep[see e.g.,][]{Wagner2015, Garufi2017}. Shadows may actually be the best vehicle to connect spirals in the outer disk and high NIR in the inner disk, in a scenario where the reduced gas pressure under the shadow may (contribute to) excite spiral waves \citep{Montesinos2016}. While in some disks like HD100453 \citep{Benisty2017} the morphological connection between shadow and spiral seems obvious, in others, like HD135344B \citep{Stolker2017}, shadows are variable in both azimuthal position and contrast, leaving the association to spirals less straightforward.

An alternative view is that a planetary companion within the disk cavity is directly responsible for the spiral wave excitation in the outer disk \citep[e.g.,][]{Ogilvie2002} and at the same time for the inner disk misalignment. In this scenario, spirals and shadows would have the same origin but different physical mechanisms. On the other hand, gravitational instability \citep[e.g.,][]{Durisen2007} is an unlikely origin for spirals since ($i$) it requires massive disks whereas Spiral disks have typically low mass (see Fig.\,\ref{Dust_time}a), ($ii$) it is more efficient at earlier evolutionary stages while Spiral disks are old (see Sect.\,\ref{Discussion_timescale}), and ($iii$) it does not explain the high NIR of these sources. In any case, the old age of Spiral disks implies that the mechanisms responsible for the spiral formation occur or become dominant only later on in the disk/stellar evolution. In the planetary scenario, this is speculatively explained by the fact that the disk breaking necessary for the formation of a misaligned inner disk is favored when the disk aspect ratio and temperature are low \citep{Facchini2018}.

It must be noted at this stage that these considerations only apply to the quasi-symmetric, bright spirals detected in PDI on a small scale (10$-$30 au). Very extended, tenuous spirals on a larger scale, like in  HD100546 for example, and spirals detected with ALMA \citep[like Elias 2-27,][]{Perez2016} likely have a different origin.

\subsubsection{Rings} \label{Discussion_rings}
Finally, our study does not highlight any peculiar property associated with disks that have rings. In fact, the 12 disks of this type are never associated to any outer stellar companion, their NIR and FIR excess is moderate, they never show any shadow in PDI, and their hosting star is never particularly variable. Instead, their dust mass spans the almost entire interval shown by our sample. All in all, the perspective of this work is that disks with rings are the normality and our analysis cannot disentangle between their origins, being for example the interaction with companions \citep[e.g.,][]{vanBoekel2017} or the accumulation and growth of dusty material at the various ice-lines \citep[e.g.,][]{Zhang2015}.

\subsection{Faint disks in scattered light} \label{Discussion_faintness}
\begin{figure} 
  \centering
 \includegraphics[width=9cm]{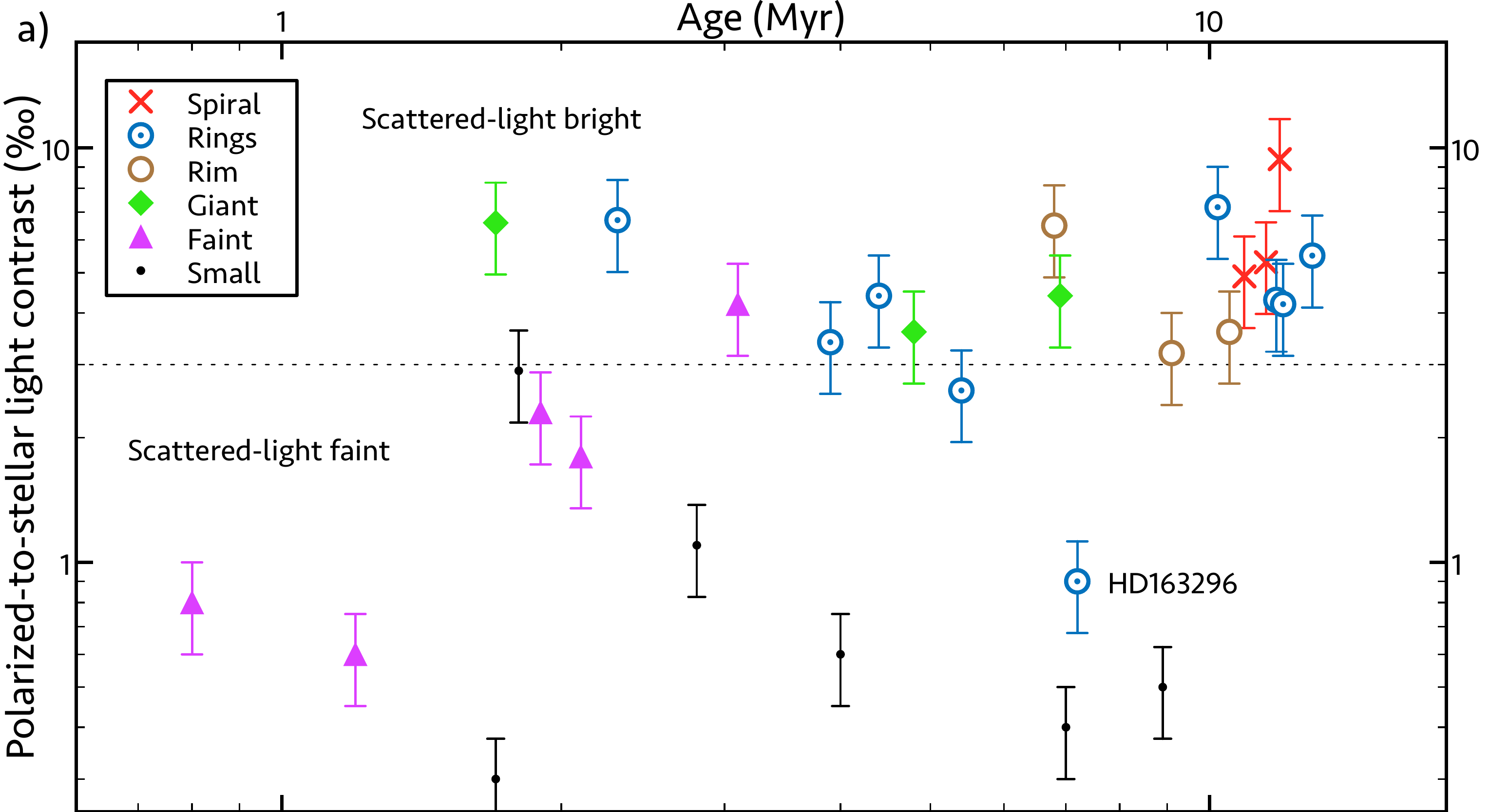}
 \includegraphics[width=9cm]{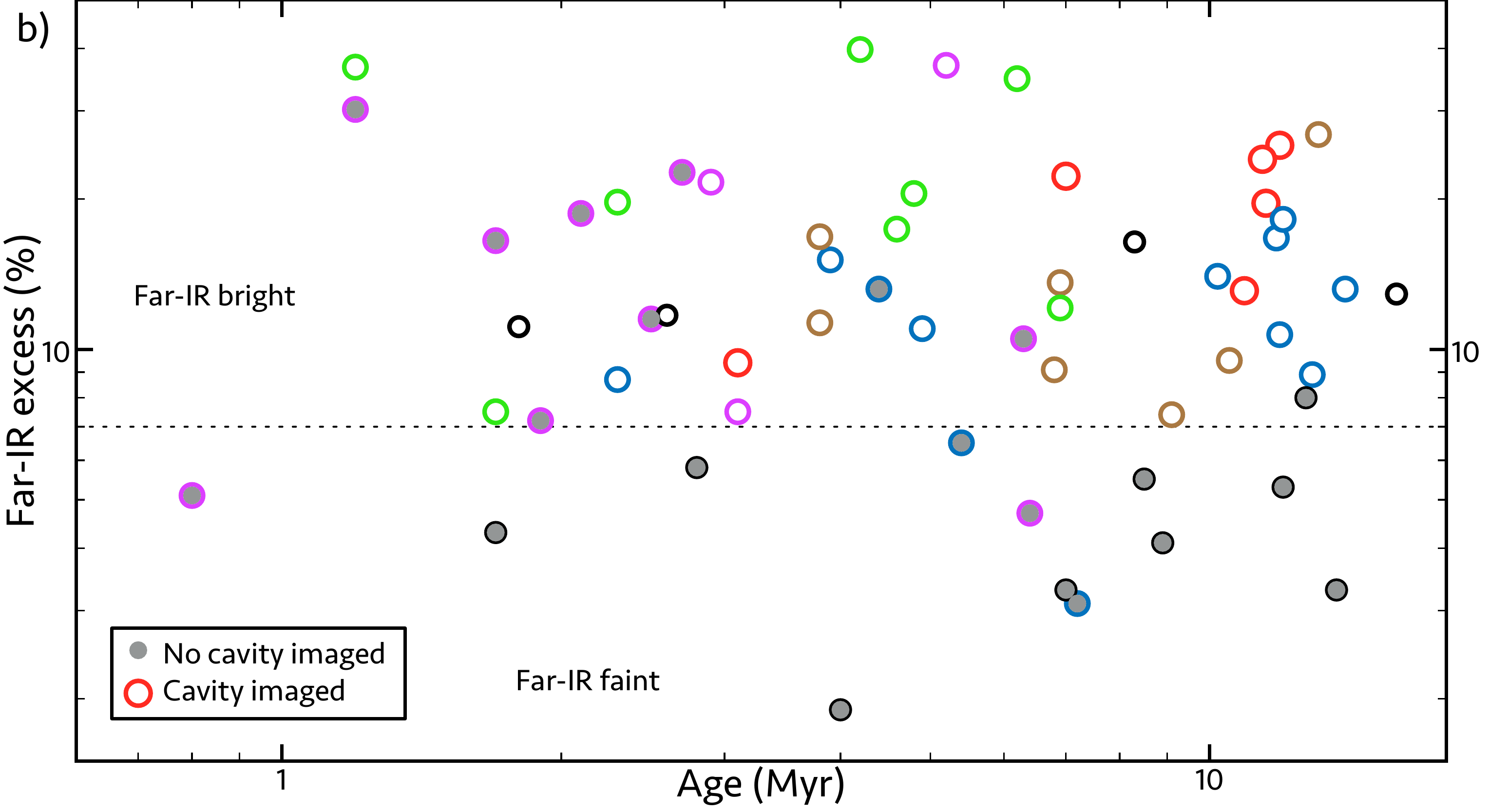}
   \caption{Temporal evolution of the polarized light contrast (a) and FIR excess (b) for some illustrative sources, in (a), and the entire sample, in (b). The values for the contrast are from \citet{Garufi2017} and \citet{Avenhaus2018}.}
 \label{FIR_contrast_time}
 \end{figure}

It is known that protoplanetary disks appear generally bright in the visible/NIR scattered light. However, up to $\sim$25\% of our disks appear faint. The majority of these are found around young stars. In other words, the amount of scattered light generally tends to increase with the stellar age, as can be seen from Fig.\,\ref{FIR_contrast_time}a for some illustrative sources. Broadly speaking, causes for the disk faintness can be diverse. HD179218 and DZ Cha have low dust mass (see Fig.\,\ref{Dust_time}a) despite being young. This is possibly explained by the presence of an outer companion \citep{Thomas2007} and photoevaporation \citep{Canovas2017}, respectively.

HD163296 and HD31648 are the prototypical Group II objects, namely those Herbig stars with a moderate FIR excess that is fitted by a single power-law continuum \citep{Meeus2001}. Historically, these objects were considered as an evolved, flattened version of the flared Group I disks. Nowadays, the properties of Group II, including faintness in scattered light, are best explained by the absence of a disk cavity in a geometry where the outer disk remains mostly self-shadowed \citep{MuroArena2018}. As shown in Fig.\,\ref{NIR_FIR} and discussed by \citet{Garufi2017}, different disk morphologies result in the observational criteria defining Group II (mostly a FIR excess lower than 5$-$6\%): Small disks, Faint disks, Inclined disks and one Ring disk (HD163296). What all these disks have in common is the absence of a large cavity. 

Finally, the faintness of massive disks like Sz71, CI Tau, AS209, and RU Lup is more controversial since these objects have a large FIR excess (they are, formally speaking, \mbox{Group I}). A likely explanation for this contradiction is the different origin of the FIR excess; in evolved sources ($\gtrsim$3 Myr) this component is primarily originated in the directly illuminated outer disk surface, whereas in younger sources it is partly due to some uplifted material or envelope that is inherited from previous evolutionary stages. This dichotomy is clear in Fig.\,\ref{FIR_contrast_time}b, where we show that, among the oldest sources, only disks with a cavity are bright in the FIR whereas, among the youngest, this excess can be large even in disks with no cavity. Thus, in disks with no cavity the FIR excess tends to decrease with time.
The large FIR excess of many young sources does not have a counterpart in scattered light and the disk therefore looks faint. In other words, the absence of a cavity is the most likely reason why these disks are faint in scattered light even though an additional cold component unrelated to the disk surface distinguishes them from the observationally defined Group II. 

Summarizing, the disk faintness in scattered light seems to always be related to the absence of a cavity. In this framework, the accumulation of Faint disks around young stars could be proof that ($i$) some disk cavities are formed after $\approx$3 Myr, or that ($ii$) mainly disks with a cavity live for $\gtrsim$5 Myr (see Sect.\,\ref{Discussion_evolution}).

\subsection{Disk evolution and dissipation} \label{Discussion_evolution}
Our sample shows a shallow, declining trend for the dusty disk mass with time (Fig.\,\ref{Dust_time}). In principle, the trend suggests that the disk evolution is universal since our sources belong to a dozen SFRs. We must nonetheless keep in mind the scatter and the biases of the trend. Large scatter is found already in the correlation between disk and stellar mass \citep[see e.g.,][]{Pascucci2016}. Since our diagrams show the dust masses normalized to the stellar mass, this scatter is included in the trend. It is possible that this dispersion is the result of different initial conditions or evolutionary paths of protoplanetary disks. The fact that the correlation with the age is less dispersed when accounting for the different PMS timescales (see Fig.\,\ref{Dust_time}c) is a likely indication that disks around different stars evolve differently.

We showed that the current dataset of disks in scattered light represents the upper tail {of dust mass} distribution of all disks \citep[see e.g.,][for comparison]{Ansdell2016, Pinilla2018}. In particular, our sample likely represents the vast majority of the most massive protoplanetary disks within 200 pc. This is clearly an observational bias but, as such, it may be used to infer that the dissipation of material in massive disks is a slow process (as dictated by the shallow trend). Speculatively, this is explained by the propensity of more massive disks to form massive planets \citep{Mordasini2012b} and by the efficiency of these giant planets to retain part of the dusty material at large disk radii for a longer time \citep{Pinilla2012}.   

This scenario may also reconcile with the age inhomogeneity of our sample. In fact, the number of disks older than 10 Myr is comparable to that of disks of 1$-$3 Myr, whereas this ratio is 1:7 based on the excess at 20 $\mu$m of large surveys \citep[see][and references therein]{Ercolano2017}. Interestingly, the generally reported disk lifetime of 4$-$6 Myr \citep[see e.g.,][]{Ribas2014} approximately corresponds to the age of the oldest Faint (cavity-less) disks of our sample. After this time, all the non-inclined disks show a cavity in the (sub-)millimeter and are bright in scattered light. One potential explanation for the high occurrence of transition disks among the old sources is the aforementioned propensity to form massive planets. Thus, massive disks with large cavities may have different evolutionary paths from the others \citep[see e.g.,][]{Pinilla2018} and the disk features that we observe, particularly the spirals, may be less common in short-lived, cavity-less disks.  

As for the less massive disks that are not included in this sample, it is possible that their evolution is far more rapid. As mentioned in Sect.\,\ref{Discussion_faintness}, the low mass of the outliers HD179218 and DZ Cha is best explained by the tidal interaction with an outer companion and internal photoevaporation, respectively. According to \citet{Long2018}, these two mechanisms are the most likely responsible for the low-mass disks that do not follow the disk-stellar-mass scaling law. As for the former explanation, the actual number of small disks (possibly truncated by an outer companion) is much larger than that of this sample \citep[see e.g.,][]{Barenfeld2017}. Also, the fraction of stellar systems in our sample ($\approx$20\%) is much lower than the real one \citep[$\approx$50\%,][]{Duchene2013}, suggesting the existence of many low-mass disks that would steepen the temporal decline of the disk mass in Fig.\,\ref{Dust_time}.

\subsection{Current and future framework of PDI observations}
Imaging protoplanetary disks with PDI in the NIR has two important limitations. First, the angular resolution (although excellent) and small size of protoplanetary disks limit the observability range to a distance smaller than 400 pc (preferably even smaller than 200 pc). Second, the adaptive optics at the telescope set a certain threshold in the apparent optical or NIR brightness of the star (dependent on the instrument). Both a low intrinsic luminosity and a high circum- and interstellar extinction contribute to lowering the apparent magnitude. These limitations concur to confine the available sample to the stellar masses and ages shown in Fig.\,\ref{Star_properties}c {and Table \ref{Limit_Rmag}. To quantify the impact of these limitations on the current sample, we derived the fraction of stars from Taurus \citep[from][]{Andrews2013}, Lupus \citep{Ansdell2016}, Chamaeleon I \citep{Long2017}, and Upper Sco \citep{Barenfeld2017} with R mag < 13. This yielded 51\%, 31\%, 30\%, and 46\%, respectively. Therefore, we are currently only able to access less than half of the PMS stars within 200 pc.} 

PDI observations of high-mass stars ($M>3\, {\rm M_{\odot}}$) are uncommon because of the scarcity of nearby O and early-B stars, while those of very low-mass stars ($M<0.5\, {\rm M_{\odot}}$) are challenged by their low brightness. This element also introduces a bias to the age of moderately low-mass stars ($M<1\, {\rm M_{\odot}}$), as only those younger than 5$-$6 Myr are bright enough. In this work, we also showed that all disks younger than 2$-$3 Myr are difficult targets in PDI because of both ($i$) the extended emission occulting the star or dominating the polarimetric contribution and ($ii$) the paucity of large cavities that leaves the outer disk substantially under illuminated.

All in all, PDI applied to this generation of telescopes has been very useful for the detection of disk features around single, relatively evolved, intermediate-mass stars \citep[like e.g., HD135344B and HD169142,][]{Muto2012, Quanz2012} or exceptionally extended and bright objects \citep[like GG Tau and IM Lup,][]{Yang2017, Avenhaus2018} but the characterization of brown-dwarf disks, small disks, and early-stage disks will be hindered by the aforementioned limitations. Nevertheless,  enlarging the sample to more TTSs \citep[proceeding e.g., the line of investigation by][]{Avenhaus2018} is of pivotal importance to confirm or reject the trends highlighted in this work.

\begin{figure} 
  \centering
 \includegraphics[width=9cm]{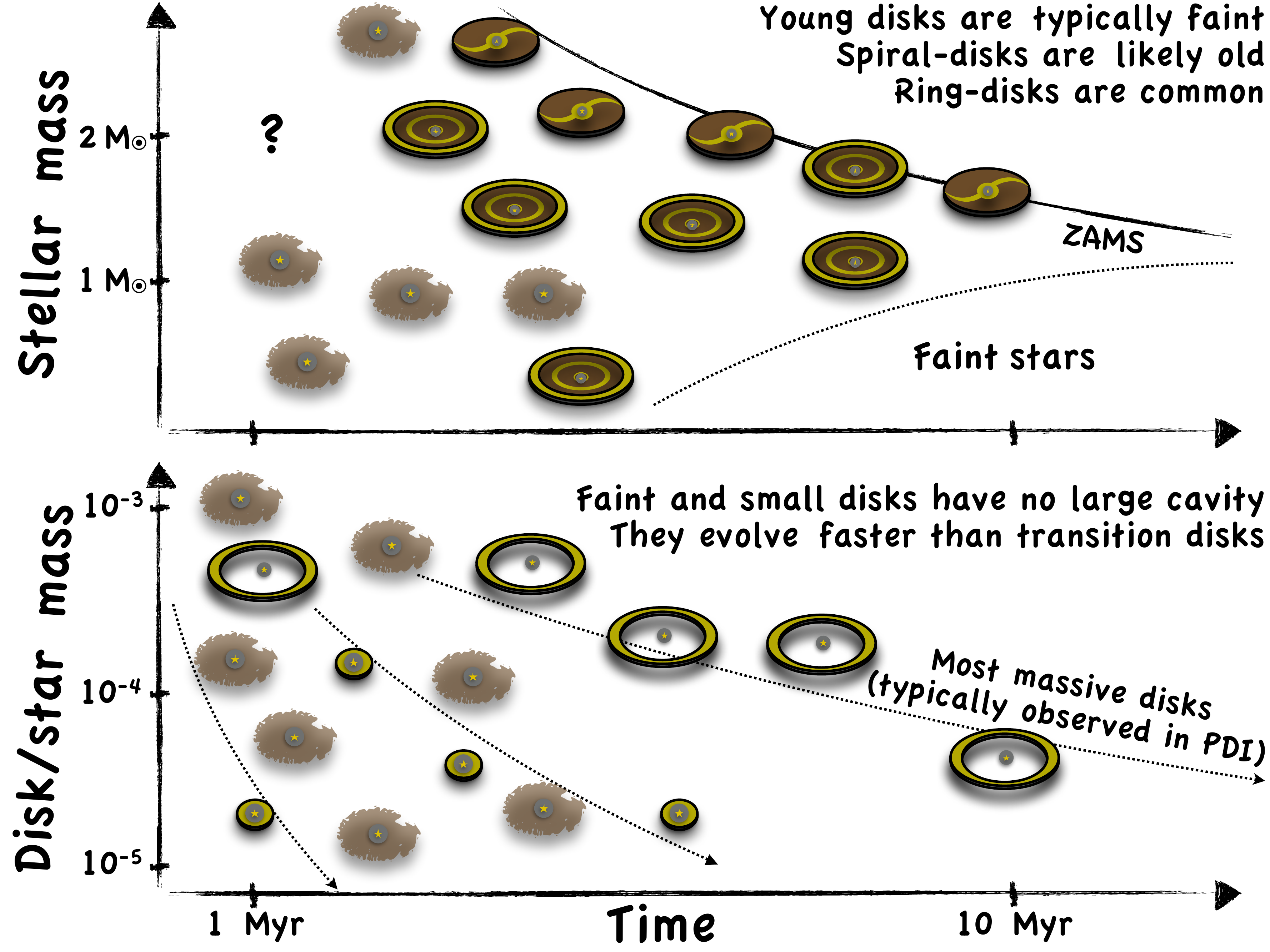}
   \caption{Sketch summarizing the results on the taxonomical analysis of 58 protoplanetary disks in scattered light. {The above panel emphasizes the young age of faint disks and the old age of spiral disks relative to the respective PMS stellar lifetime. It also highlights the space of stellar properties that is not covered by the currently available sample. The below panel sketches the shallow trend of the dust mass with time for the transition disks that are typically observed in PDI.}}
 \label{Final_sketch}
 \end{figure}

\section{Conclusions} \label{Conclusions}
In this work, we investigate the origin and the timescale of protoplanetary disk substructures through a taxonomical analysis of a large set of scattered-light images from the literature. We classify 58 disks into six major categories based on their appearance in polarimetric light (PDI). The comparison with their newly calculated stellar and disk properties from GAIA DR2 has led to the following major trends.
\begin{enumerate}
\item Disk substructures are always seen when the disk is extended and bright enough (given the resolution and contrast).
\item Young disks ($\lesssim$5 Myr) are typically faint in scattered light, whereas old disks are systematically bright. 
\item Disks observed thus far show a loose, shallow decline of dust mass with time. The outliers are typically truncated disks.
\item The presence of spirals and shadows is associated to a high NIR excess. Disks with rings have low NIR excess.
\item Spirals are found only around stars toward the end of their PMS evolution ($\gtrsim$80\% of the PMS lifetime).
\item Up to 70\% of the disk cavities is significantly smaller in scattered light than in the (sub-)millimeter.  
\end{enumerate}

We also investigated the framework of the current dataset, revealing that primarily evolved super-solar and young sub-solar mass stars have been observed thus far. The former limit may be related to the embedding envelope of early-type, massive stars. The latter is instead due to the low brightness of evolved TTSs, which is technically challenging for the current instruments. We also found a relatively strong bias toward  single stars,  massive disks, and transition disks.

Our conclusions based on points (1) and (2) are that disks may always host substructures but these remain undetected in small disks (10$-$20 au in size) and in several young disks, since they are faint. Their faintness is likely related to the absence of a disk cavity, in a scenario where most disks older than $\sim$5 Myr host a cavity, and are therefore bright. The general preference to observe this type of object explains why the disk mass evolution is so flat (point 3), since the bulk of the disk population (composed of small disks in stellar systems and rapidly evolving disks) remains mostly unobserved in scattered light. In other words, the current PDI dataset includes a large fraction of the disks with a lifetime longer than the e-folding time of $\approx$5 Myr \citep{Ribas2014}, and it therefore represents a particular path of disk evolution. This view is summarized in Fig.\,\ref{Final_sketch}.

Points (4) to (6) are all suggestive of the presence of planetary companion(s) within the cavities. In fact, a giant planet of at least $\approx$1 $M_{\rm Jup}$ may perturb the orientation and the scale height of the inner disk thus generating a high NIR and shadows on the outer disk. The same planet(s) or, indirectly, the shadows, can then trigger the spiral waves. In this scenario, the absence of spirals around TTSs may be due to ($i$) our inability to observe the late PMS evolution of sub-solar stars (see Fig.\,\ref{Final_sketch}), or to ($ii$) the inefficiency of less massive disks to form massive planets. In any case, gravitational instability as the origin of spirals is less plausible, since the disk masses of these objects are on average lower than the others. A giant planet is also the most likely explanation for the small size of NIR cavity compared to the millimeter, since planets differentially filter the dust grains. However, these planets should not be more massive than $\approx$5 $M_{\rm Jup}$ to allow the replenishment of the inner disk that is always present in our sample.

Currently only a small fraction of our sample has complementary (sub-)millimeter images with a comparable angular resolution (from ALMA). However, this fraction will certainly increase in the near future. The results of this work will therefore have to be confronted with the newer PDI and ALMA observations that the community will soon carry out.

\begin{acknowledgements}
We thank the SPHERE consortium for making some GTO unpublished data available for the paper. We acknowledge the referee for the constructive report. This research has made use of the VizieR catalogue access tool, CDS, Strasbourg, France. The original description of the VizieR service was published in A\&AS 143, 23. This work has been supported by the project PRIN-INAF 2016 The Cradle of Life - GENESIS-SKA (General Conditions in Early Planetary Systems for the rise of life with SKA). A.G.\ acknowledges the support by INAF/Frontiera through the "Progetti Premiali" funding scheme of the Italian Ministry of Education, University, and Research. We acknowledge funding from ANR of France under contract number ANR-16-CE31-0013 (Planet Forming disks). P.P.\ acknowledges support by NASA through Hubble Fellowship grant HST-HF2-51380.001-A awarded by the Space Telescope Science Institute, which is operated by the Association of Universities for Research in Astronomy, Inc., for NASA, under contract NAS 5-26555.
\end{acknowledgements}

\bibliographystyle{aa} 
\bibliography{Reference.bib} 

\longtab{
\centering
\begin{landscape}
\begin{longtable}{c|c|cccccccc}
\caption{\label{Sample_literature} Literature properties of the sample. \textit{The complete table will appear in the journal version.} Columns are: reference number in this work; main name; alternative name from the literature; star forming region; distance from GAIA DR2 \citep{Gaia2018}; effective temperature and optical extinction adopted in this work; stellar multiplicity, with star(s) indicated by asterisks and the disk by the brackets; flux at 1.3 mm, unless differently specified by the superscript number; reference for the effective temperature (letters) and millimeter flux (numbers), as specified below.}\\
\hline\hline
\# & Main & Alternative & Region & $d$ & $T_{\rm eff}$ & $A_{\rm V}$ & Multiplicity & F$_{\rm 1.3mm}$ &  Reference \\
& name & name & & (pc) & (K) & (mag) & & (mJy) & \\
\hline
\endfirsthead
\caption{continued.}\\
\hline\hline
\# & Main & Alternative & Region & $d$ & $T_{\rm eff}$ & $A_{\rm V}$ & Multiplicity & F$_{\rm 1.3mm}$ &  Reference \\
& name & name & & (pc) & (K) & (mag) & & (mJy) & \\
\hline
\endhead
\hline
\endfoot
1& HD31293 & AB Aur & Taurus & 162.9 $\pm$ 1.6  & 9800 & 0.2 & (*) & 136$\pm$15 & b,1 \\
\end{longtable}
\tablefoot{b \citet{Folsom2012}, 1 \citet{Sandell2011}.}
\end{landscape}
}

\longtab{
\centering
\begin{landscape}
\begin{longtable}{c|c|ccccccccccccccc}
\caption{\label{Sample_calculated} Properties of the sample from this work. \textit{The complete table will appear in the journal version.} Columns are: reference number in this work; main name; alternative name; stellar mass; age; age normalized by the ZAMS variability; NIR excess and FIR excess relative to the stellar flux; disk dust mass relative to the stellar mass; category of the disk as from Fig.\,\ref{Sketch}; reference for the PDI image of the disk, as specified below.}  \\
\hline\hline
\# & Main & Alternative & $M_*$ & $t$ & $t$/ZAMS & $\Delta$V & F$_{\rm NIR}$/F$_*$ & F$_{\rm FIR}$/F$_*$ & $M_{\rm dust}/M_*$ & Category & Reference \\
& name & name & ($\rm M_{\odot}$) & (Myr) & & (mag) & (\%) & (\%) & ($\cdot10^{4}$) & & \\
\hline
\endfirsthead
\caption{continued.}\\
\hline\hline
\# & Main & Alternative & $M_*$ & $t$ & $t$/ZAMS & $\Delta$V & F$_{\rm NIR}$/F$_*$ & F$_{\rm FIR}$/F$_*$ & $M_{\rm dust}/M_*$ & Category & Reference \\
& name & name & ($\rm M_{\odot}$) & (Myr) & & (mag) & (\%) & (\%) & ($\cdot10^4$) & & \\
\hline
\endhead
\hline
\endfoot
\smallskip
1& HD31293 & AB Aur & 2.3 $\pm$ 0.2 & 4.6$^{+0.9}_{-0.5}$ & 0.88 & 0.33 & 27.1 $\pm$ 2.9 & 17.4 $\pm$ 0.6 & 0.33 $\pm$ 0.03 & Giant (Spiral) & 1 \\
\end{longtable}
\tablefoot{1 \citet{Hashimoto2011}.}
\end{landscape}
}

\begin{appendix} 
\section{Caveats on the sample} \label{Appendix_sample}
The sample studied in this work is shown in Table \ref{Sample_literature}. It consists of the vast majority of sources being observed in PDI and with either published papers or works in preparation (see Table \ref{Sample_calculated} for all references). We limit the sample to protoplanetary disks, by excluding debris disks but also intermediate-stage disks (sometimes referred to as anemic disks, like e.g., HD141569). The quantifying criterion to distinguish these three categories is the FIR excess (being $0.1-1\%$ for anemic disks and $<0.1\%$ for debris disks). We also removed from the analysis those sources that are contaminated by an embedding nebulosity \citep[RY Tau and SU Aur,][]{Takami2014, Jeffers2014}. In the following, we discuss the individual sources that present either a critical classification or caveats in the properties calculation.

\begin{itemize}
\item HD142527 and GG Tau A are here considered as Giant disks but could also be labeled as Rim disks. They are certainly peculiar objects (similar to each other) with a giant cavity encompassed by a very prominent disk wall. They are both multiple stellar system, as HD142527 consists of a pair of F+M stars and GG Tau A of three M stars. 

\item HD31293, HD34282, GM Aur, IM Lup. These four Giant disks may have a controversial classification. It is possible that the multiple arms in IM Lup and GM Aur are concentric rings that appear fainter than the prototypical cases like TW Hya. In HD34282, the arm(s) can even be considered as multiple spiral arms or a unique ring. Finally, HD31293 (AB Aur) is the prototypical giant disk with multiple arms even though \citet{Tang2017} showed the existence of a double-spiral structure in CO resembling that of HD135344B.

\item HD163296 is treated as a Ring disk in this work even though the polarized signal is low (see Fig.\,\ref{FIR_contrast}). This is motivated by the fact that the ring itself is, unlike e.g., AS 209, very bright while the overall brightness is low because of self-shadowing of the inner disk regions \citep[see Sect.\,\ref{Discussion_faintness} and][]{Garufi2017}.  

\item The PDI image of DoAr28 by \citet{Rich2015} does not seem to show any peculiar structure. On the other hand, we cannot infer whether the disk is intrinsically faint. It must be noted that the disk is an outlier in the faint-young analogy, as the star is as old as 6.3 Myr.

\item In principle, a Rim disk with a significant signal from within the cavity can be considered a Ring disk where the region inside the rim is merely a disk discontinuity. PDS70 and LkCa15 are here considered Rim, while PDS66 and J1852 are considered Ring, even though their classification is possibly interchangeable.

\item The distance to HD179218 is larger than most of the other sources (266 pc). However, this discrepancy (an average factor $\approx$1.8) does not seem enough to explain the faintness of the disk by perspective considerations only. Other sources with prominently imaged disks (e.g., LkH$\alpha$330 and V1247 Ori) are much further away (310 and 398 pc, respectively). 

\item The uncertainties on all the properties of SR21 and IRS48 are much larger than the rest of the sample since the high optical extinction $A_{\rm V}$ (=6.3 and 10.0, respectively) makes the calculation of the stellar luminosity critical.

\item Unlike all other Small disks, HD144668 and DI Cha remain, to our knowledge, undetected in PDI. However, all their properties are similar to those of the other Small disks \citep[see e.g.,][]{Garufi2017} and the most likely explanation for their non-detection is the very limited radial extent. We therefore treat them here as Small disks. 
\end{itemize}

We critically assessed that the main conclusions of the paper remain unaltered by these critical cases.

\section{Observability of faint stars}
{As discussed in Sect.\,\ref{Results_stars}, the absence of old, low-mass stars observed thus far in scattered light is due to the limit in stellar brightness imposed by this generation of telescopes. Assuming a representative limit in the R band of $\sim13$ mag, this restriction translates into the stellar properties shown in Table \ref{Limit_Rmag}.}

\begin{table}
 \caption[]{Observability of young stars for some representative values assuming a lower limit to the stellar brightness in the R band of 13 mag. When the age is $<<$ 1 Myr or >20 Myr, the star is considered as never and always observable, respectively.}
 \centering
  \label{Limit_Rmag}
   \begin{tabular}{cc|cc}
    \hline
    \hline
    \noalign{\smallskip}
\multicolumn{4}{c}{Sources at 150 pc} \\
\hline
\multicolumn{2}{c|}{$A_{\rm V}=0.0$ mag} & \multicolumn{2}{c}{$A_{\rm V}=2.0$ mag} \\ 
$M_*$ (M$_{\odot}$) & Age (Myr) & $M_*$ (M$_{\odot}$) & Age (Myr) \\
\hline
0.2 & Never & 0.4 & Never \\ 
0.3 & <2 & 0.5 & <1 \\
0.4 & <3 & 0.6 & <2 \\
0.5 & <6 & 0.7 & <3 \\
0.6 & Always & 0.8 & Always \\
\hline
 \noalign{\smallskip}
\multicolumn{4}{c}{Sources at 300 pc} \\
\hline
\multicolumn{2}{c|}{$A_{\rm V}=0.0$ mag} & \multicolumn{2}{c}{$A_{\rm V}=2.0$ mag} \\ 
$M_*$ (M$_{\odot}$) & Age (Myr) & $M_*$ (M$_{\odot}$) & Age (Myr) \\
\hline
0.4 & Never & 0.9 & Never \\
0.5 & <1 & 1.0 & <2 \\
0.6 & <2 & 1.1 & Always \\
0.8 & <6 &  & \\
0.9 & Always & & \\    
    \hline
    \hline
  \end{tabular}
\end{table}

\section{Stellar variability}
\begin{figure*}
  \centering
 \includegraphics[width=4.5cm]{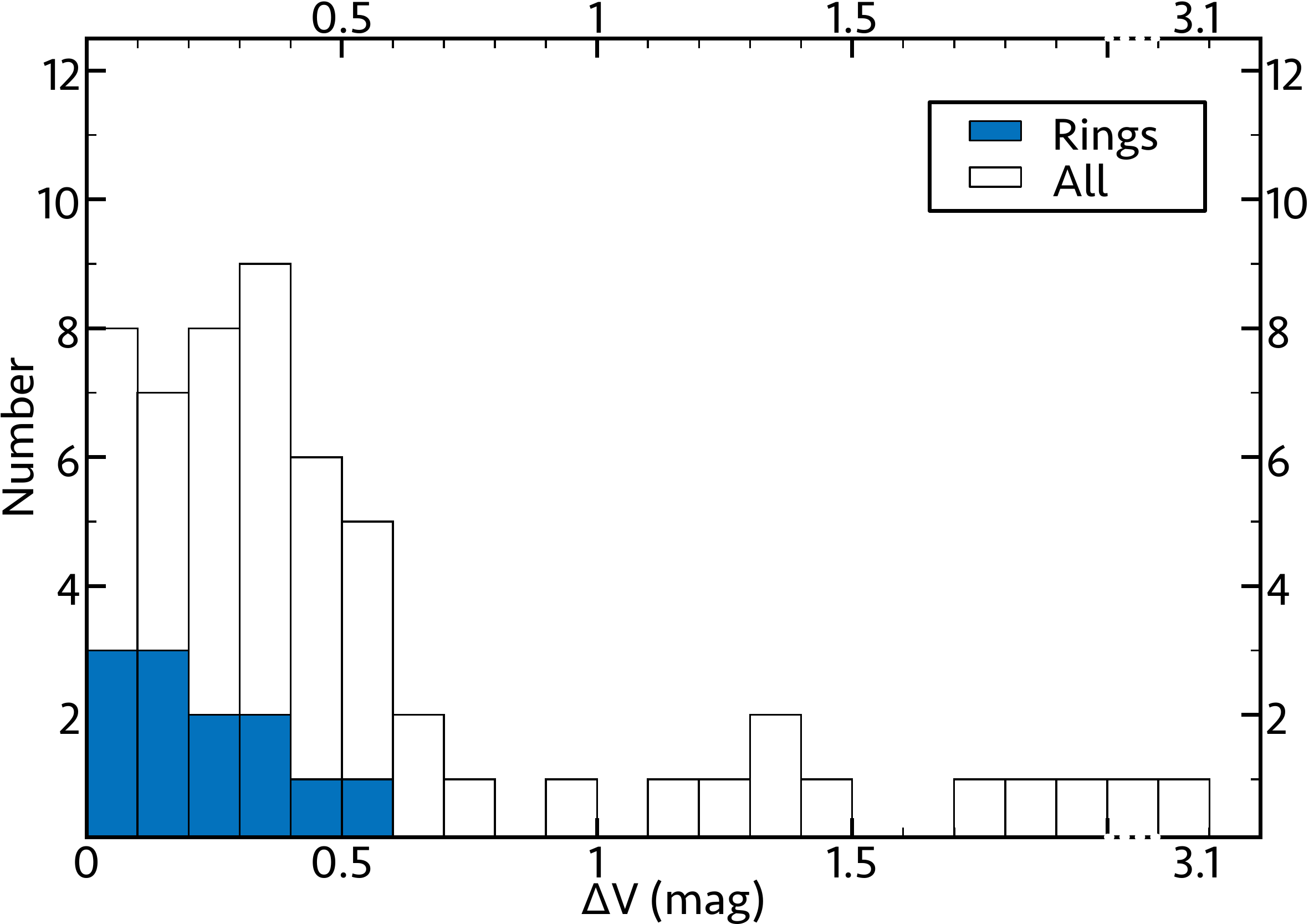}
 \includegraphics[width=4.5cm]{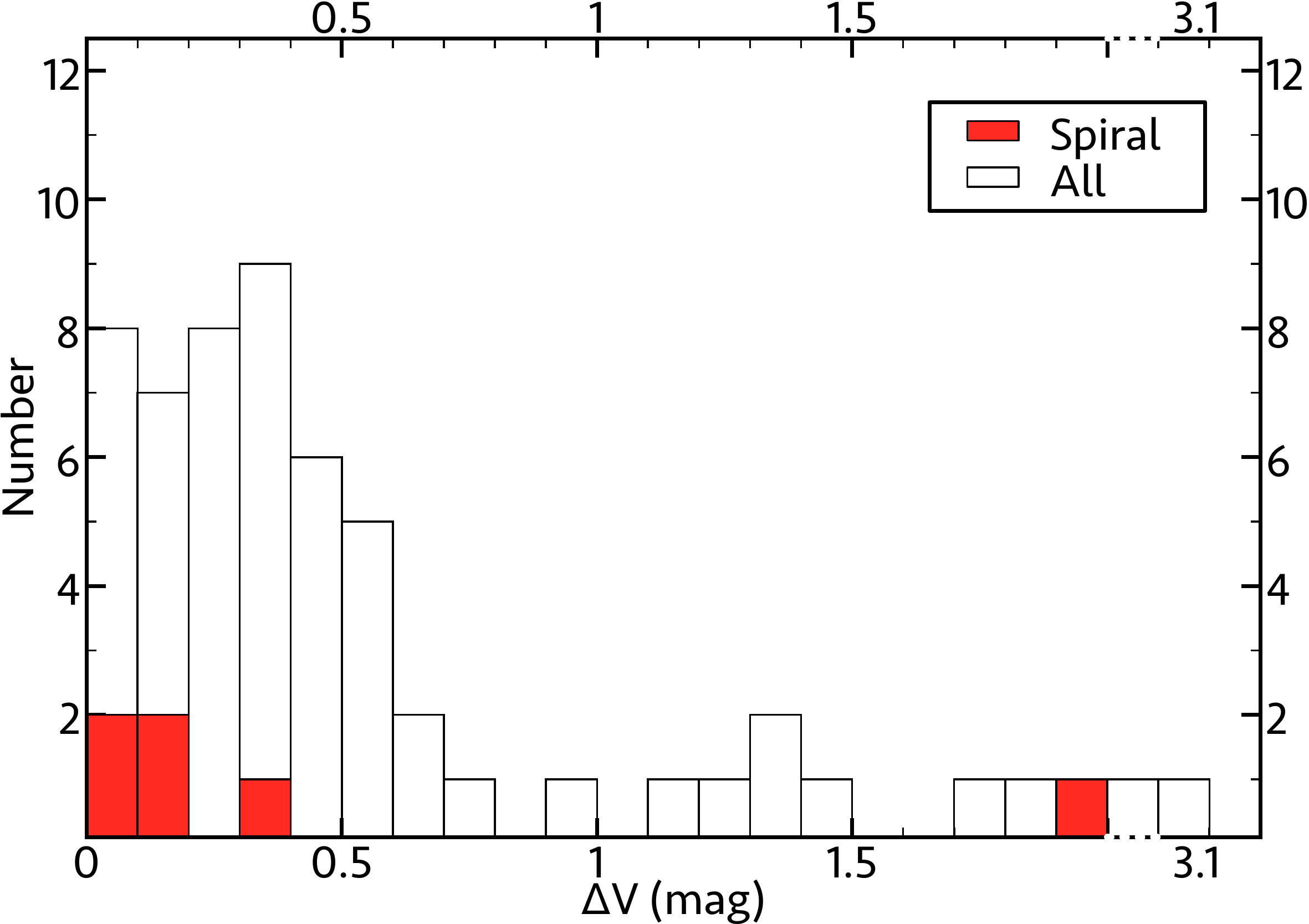}
 \includegraphics[width=4.5cm]{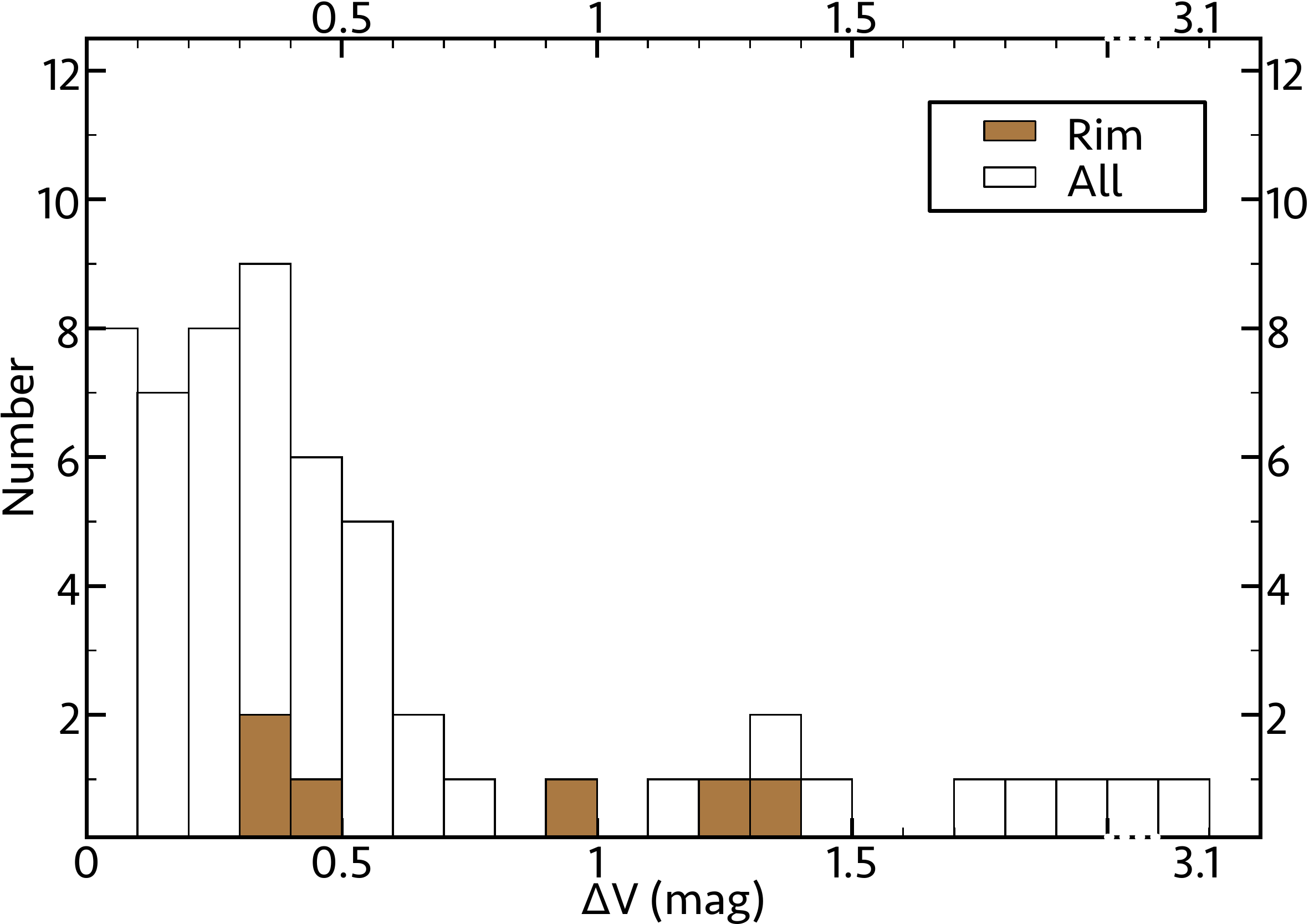}
 \includegraphics[width=4.5cm]{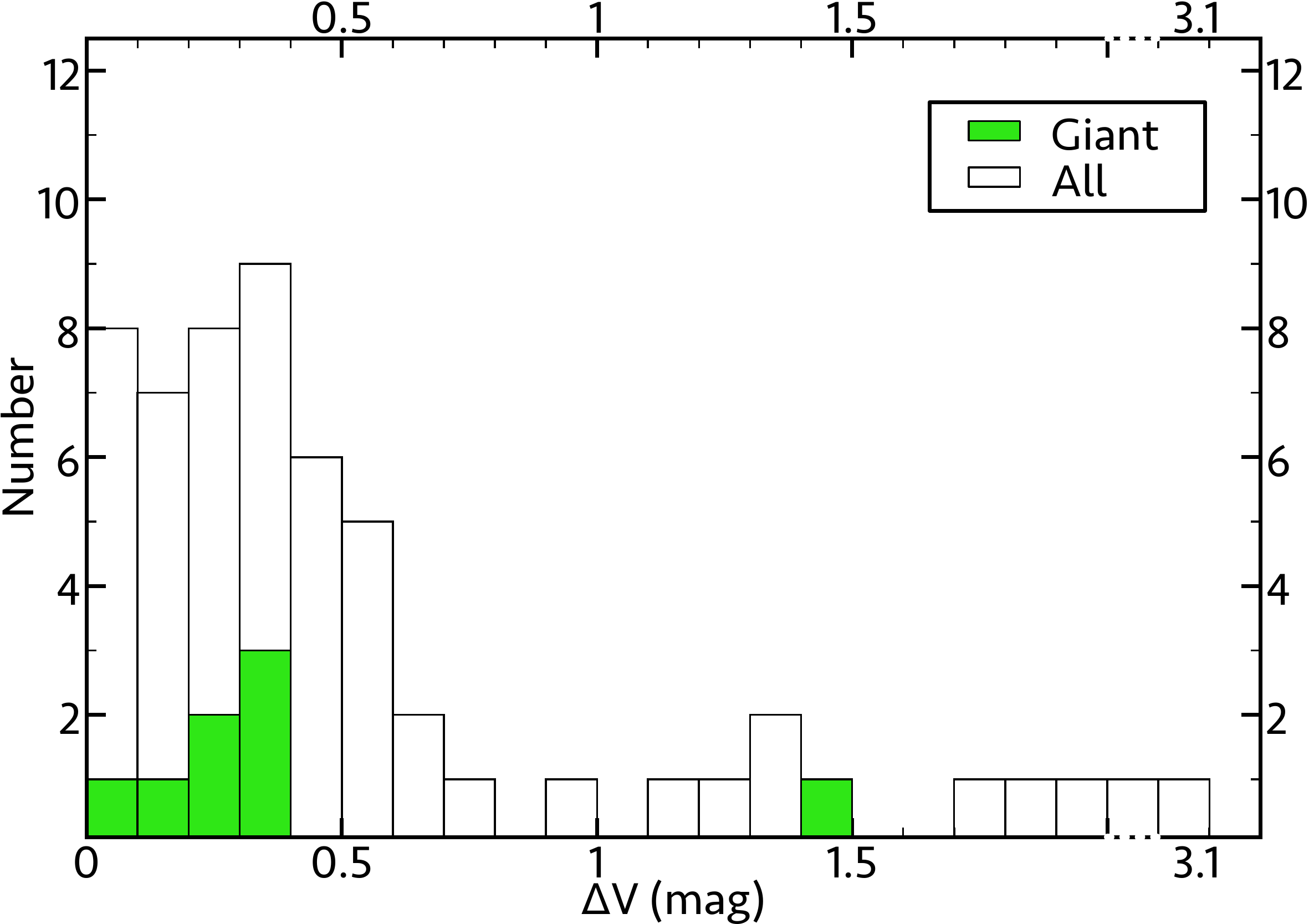}
 \includegraphics[width=4.5cm]{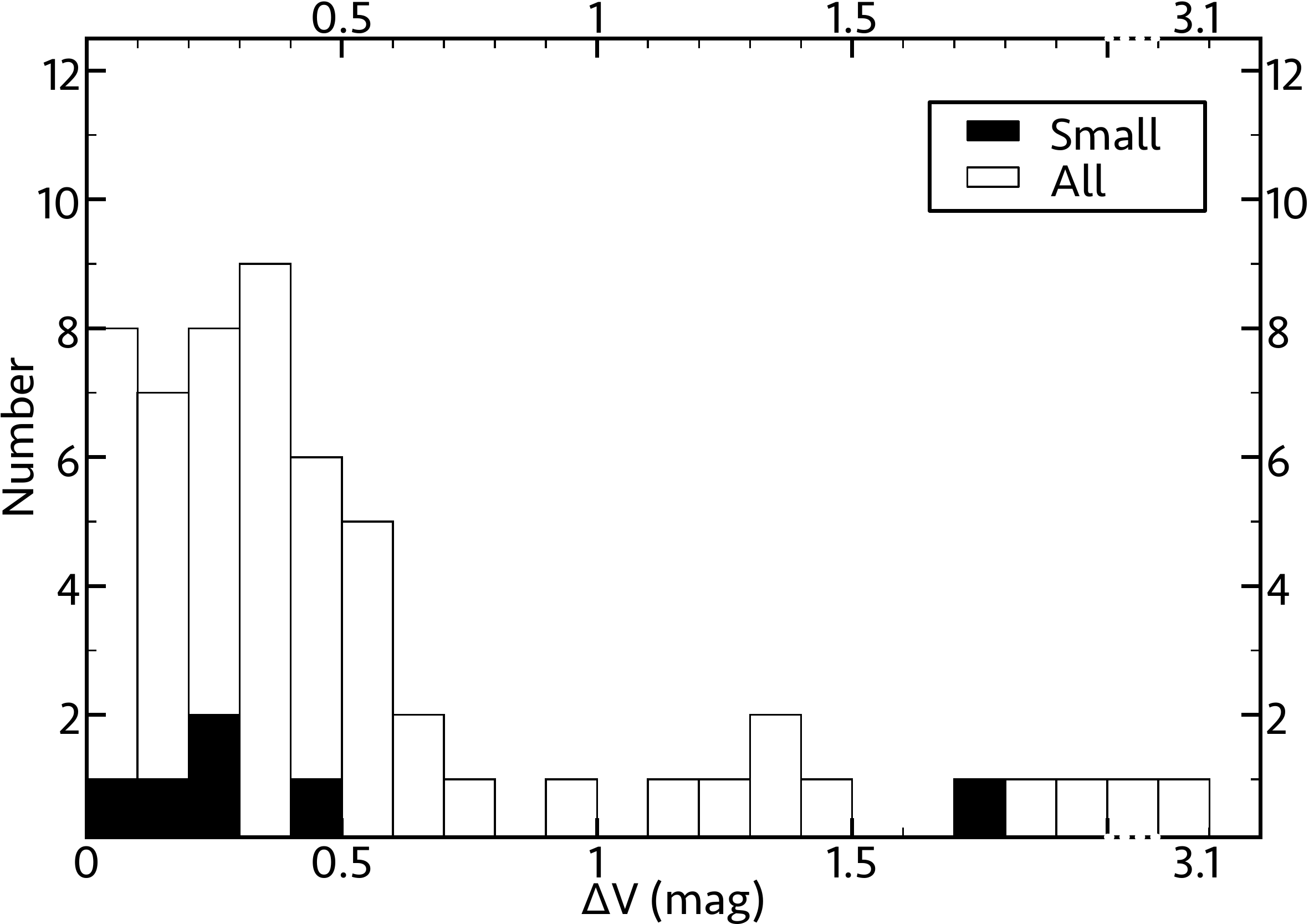}
 \includegraphics[width=4.5cm]{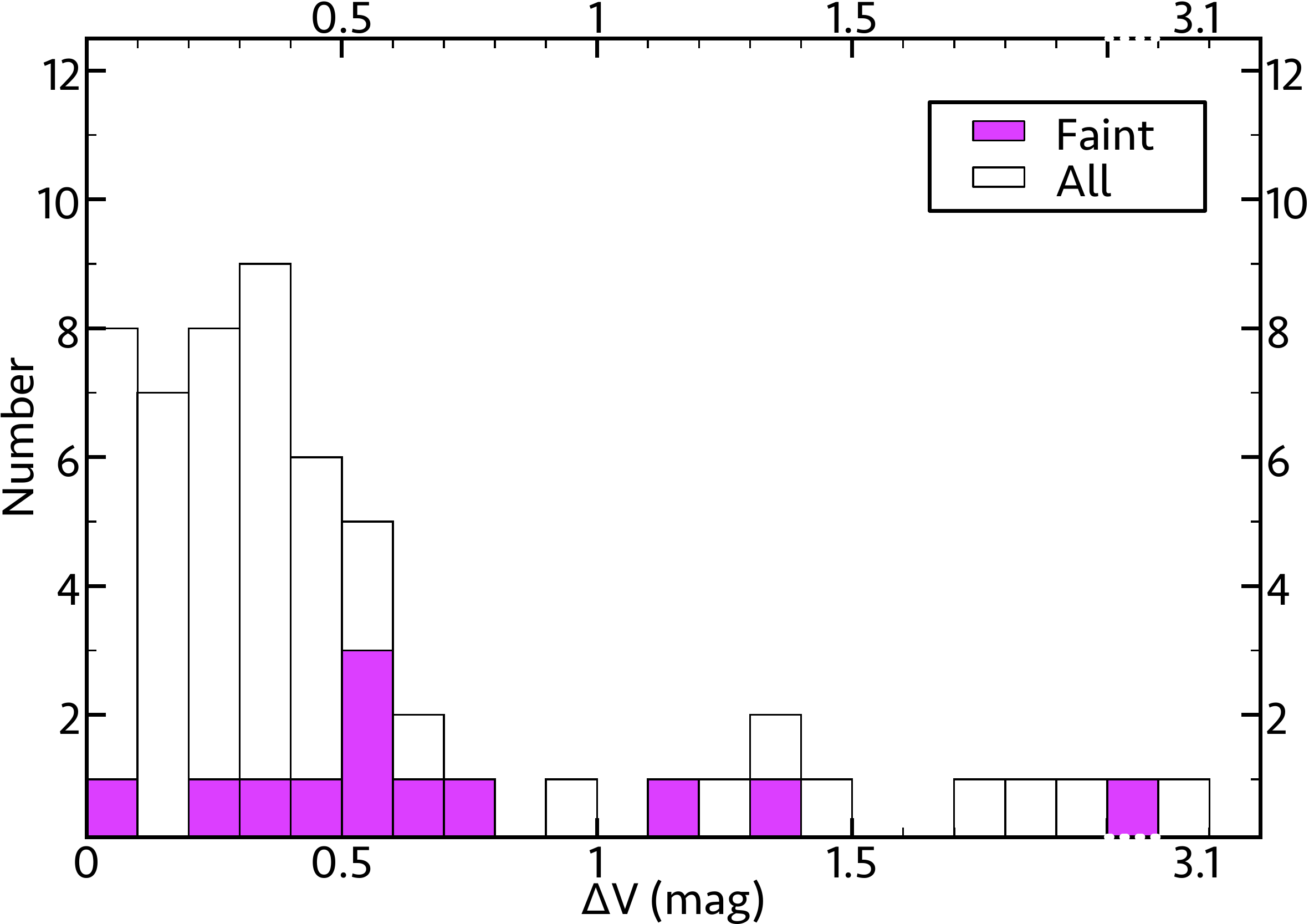}
 \includegraphics[width=4.5cm]{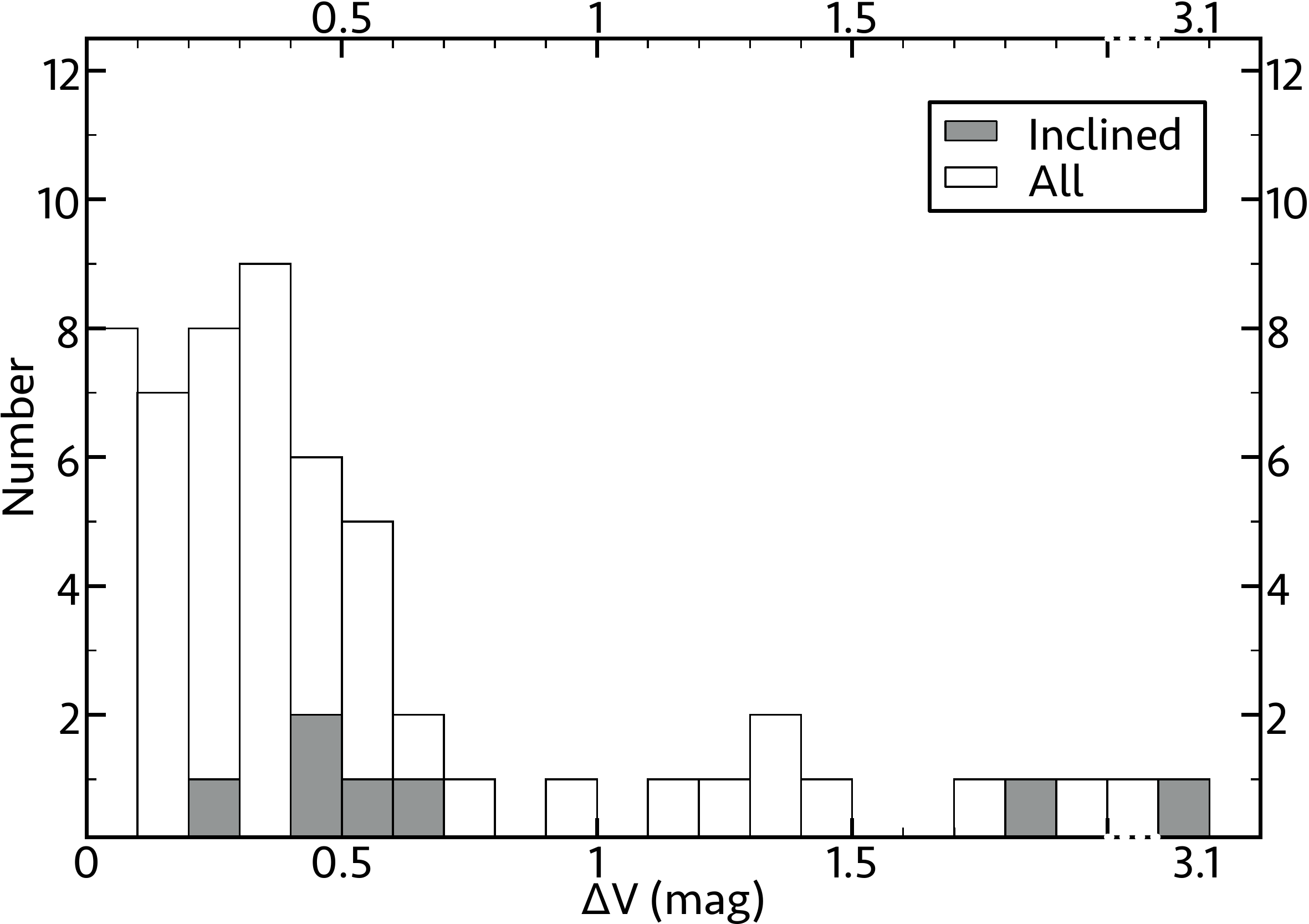}
 \includegraphics[width=4.5cm]{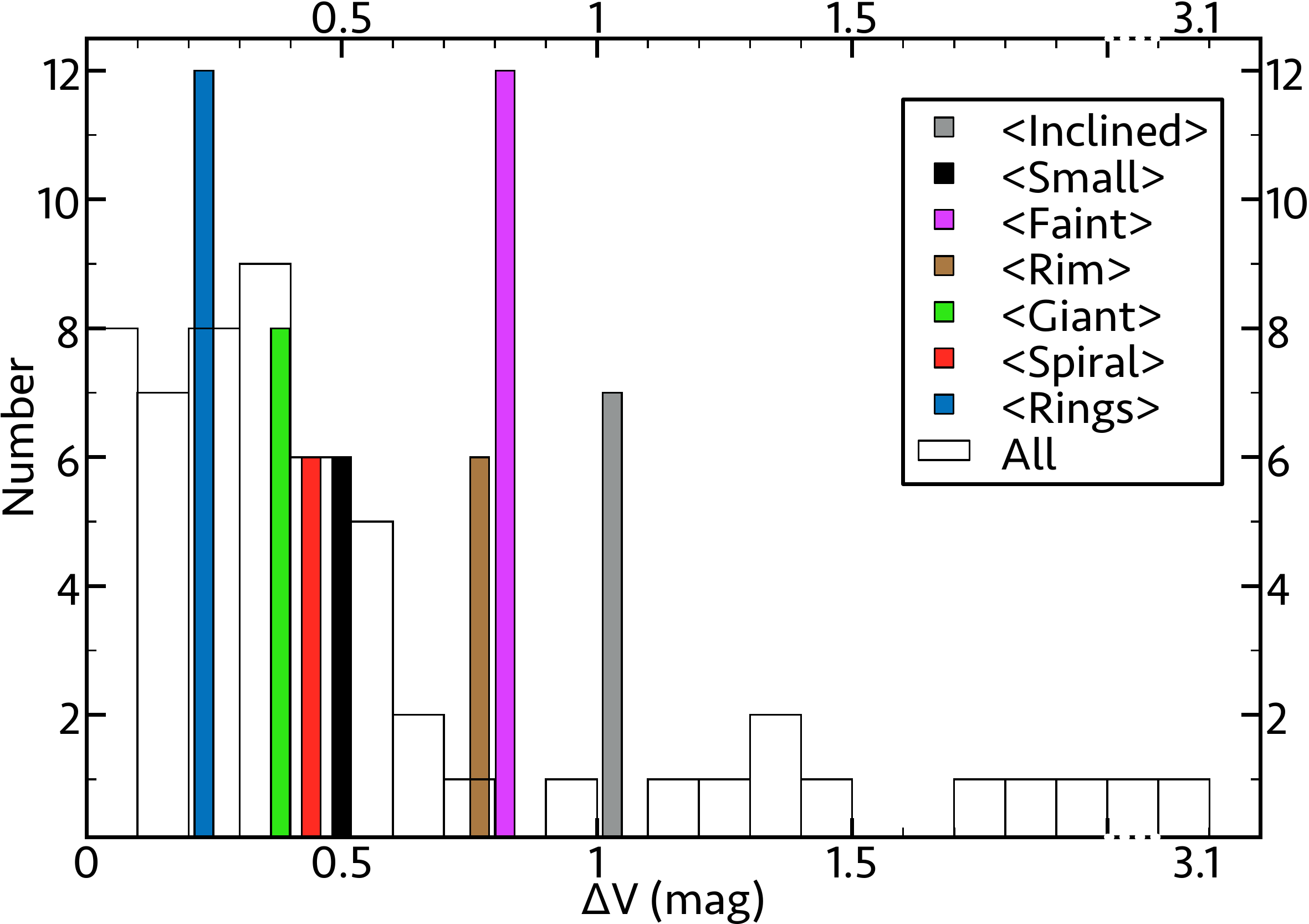}
   \caption{Stellar variability $\Delta$V of the entire sample as calculated in Sect.\,\ref{Recalculation_stars}. The narrow bars of the last box indicate the average of the various categories and their height the number of objects contributing to the average. Note the discontinuity in all boxes at $\Delta$V=2.}
 \label{Variability}
 \end{figure*}

The stellar variability $\Delta$V calculated in Sect.\,\ref{Recalculation_stars} for all the sources of this work is shown in Fig.\,\ref{Variability}. As commented in Sect.\,\ref{Results_stars}, stars with a Ring disks appear less variable than other categories like the Faint disks or the Inclined disks (being their average 0.2, 0.8, and 1.0 respectively). The number of objects in each category concurring to the average is typically low but in the case of the Ring and Faint categories is reasonably high (12) to infer that this is a significant finding. Investigating the origin of the variability for each source is worthwhile but we defer this analysis to a dedicated work.

\section{Faint disks} \label{Appendix_faint}
\begin{figure}
  \centering
 \includegraphics[width=9cm]{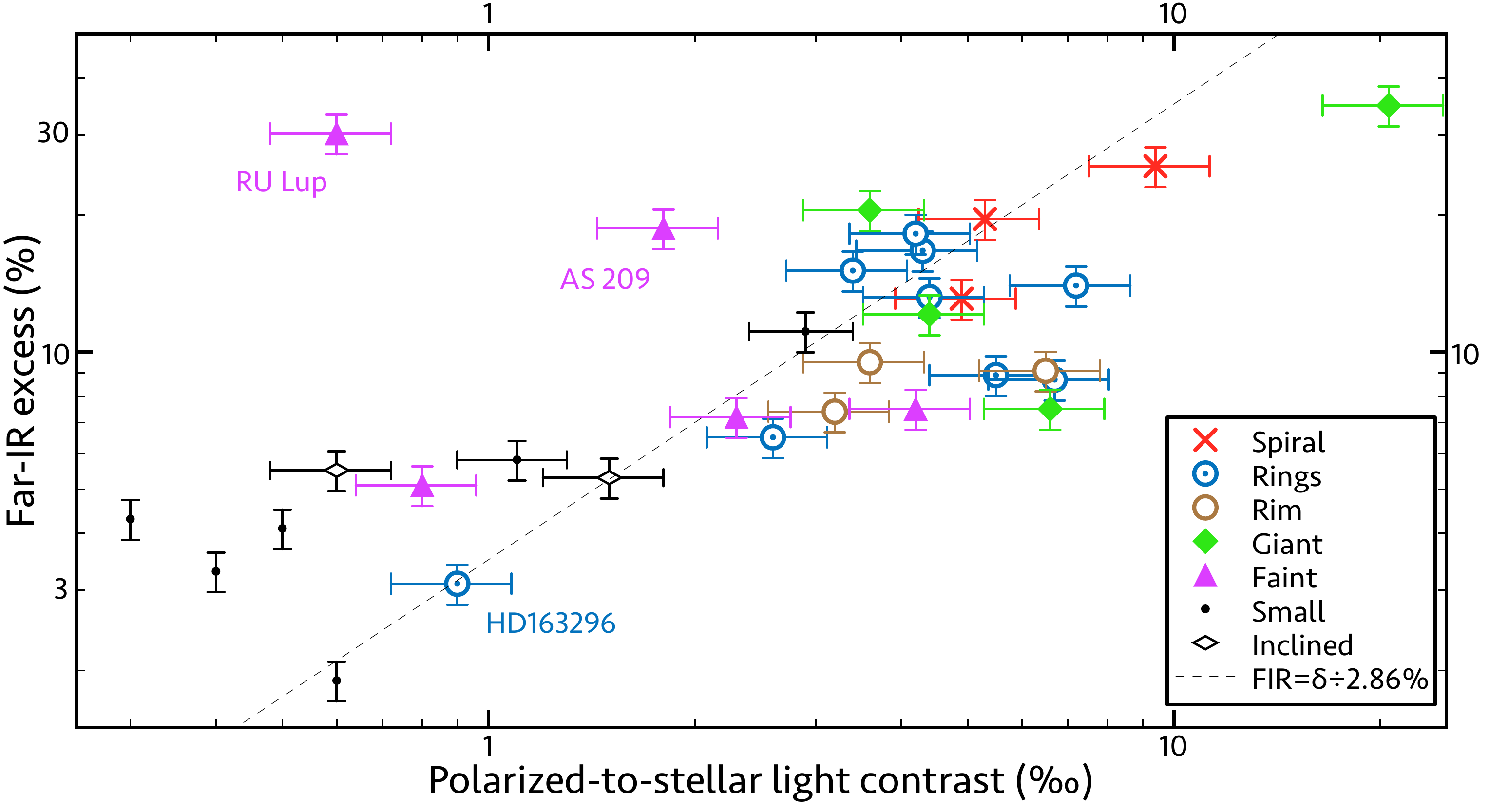}
   \caption{FIR excess vs polarized-to-stellar light contrast for some illustrative sources \citep[][]{Garufi2017, Avenhaus2018}. The dashed line is the best fit extracted by \citet{Garufi2017} for their sample.}
 \label{FIR_contrast}
 \end{figure}

What we define Faint disks in this work are sources with an actually low amount of scattered light, as can be seen in Fig.\,\ref{FIR_contrast} for some illustrative sources. This figure is an extension to that presented by \citet{Garufi2017}. Beside the Faint disks, other sources with low scattered-to-stellar light contrast are the Small and the Inclined disks, as well as HD163296 (see Appendix \ref{Appendix_sample}). From the figure, it is also clear that RU Lup and AS 209 are clearly outliers to the trend, having a prominent FIR excess but low scattered-light brightness \citep[see][and Sect.\,\ref{Discussion_faintness}]{Avenhaus2018}. 

\end{appendix}

\end{document}